\documentclass[
 reprint,
 amsmath,amssymb,
 aps,
prl
]{revtex4-2}

\usepackage{graphicx}% Include figure files
\usepackage{ragged2e}
\usepackage{subcaption}
\usepackage{dcolumn}% Align table columns on decimal point
\usepackage{bm}% bold math
\usepackage{orcidlink}

\newcommand{\ket}[1]{|#1\rangle}
\newcommand{\bra}[1]{\langle#1|}
\newcommand{\Ket}[1]{\left|#1\right\rangle}

\newcommand{\braket}[2]{\langle#1|#2\rangle}
\newcommand{\cg}[3]{\langle#1,#2;#3\rangle}

\newcommand{\Tr}{\mathrm{Tr}}

\DeclareCaptionJustification{justified}{\justifying}
\begin{document}

\preprint{APS/123-QED}

\title{Asymptotic teleportation schemes bridging between standard and port-based teleportation}
\author{Ha Eum Kim$^{1,*}$ \orcidlink{0009-0004-2614-0834}}
\email{hekim007@korea.ac.kr}

\author{Kabgyun Jeong$^{2,3,*}$ \orcidlink{0000-0001-7628-7835}}
\email{kgjeong6@snu.ac.kr}

\affiliation{%
$^1$ Department of Physics, Korea University, Seoul 02841, Korea\\
$^2$ Research Institute of Mathematics, Seoul National University, Seoul 08826, Korea\\
$^3$ School of Computational Sciences, Korea Institute for Advanced Study, Seoul 02455, Korea
}%

\date{\today}% It is always \today, today,
             
\begin{abstract}
Various modified quantum teleportation schemes are proposed to overcome experimental constraints or to meet specific application requirements for quantum communication.
Hence, most schemes are developed and studied with unique methodologies, each with its inherent challenges.
Our research focuses on interconnecting these schemes appearing to be unrelated to each other, based on the idea that the unique advantages of one scheme can compensate for the limitations of another. 
In this paper, we introduce an asymptotic teleportation scheme requiring the receiver to perform a classical selection task followed by a quantum correction.
This scheme bridges standard teleportation with port-based teleportation through the transformation of joint measurements.
Specifically, we categorize and analytically investigate protocols within this scheme for qubit systems.
Given that linear optics teleportation protocol without ancilla qubits is contained in the two non-trivial groups, we provide a novel perspective on its expansion.
Furthermore, we discuss the potential application of a protocol from one of these groups as a universal programmable processor and extend these protocols to higher-dimensional systems while maintaining the same properties and potential, providing the analytic form of the joint measurement and its performance.
These results thereby propose new avenues for developing a quantum network in higher-dimensional systems.
\end{abstract}

\maketitle

\section{Introduction}

Quantum teleportation, initially proposed by Bennett \textit{et al.} \cite{bennett1993teleporting} and also known as standard teleportation (ST), represents one of the most intriguing predictions of quantum mechanics, allowing for the transmission of an unknown quantum state across spatially separated locations without the physical transfer of particles.
This paradigm-shifting protocol exploits the non-local properties of quantum entanglement, a phenomenon that Albert Einstein famously critiqued as ``spooky action at a distance."
Over the past three decades, quantum teleportation has not only been theoretically refined but also experimentally demonstrated \cite{bouwmeester1997experimental,boschi1998experimental,furusawa1998unconditional} and developed \cite{Valivarthi2016quantum,Ren2017ground,Barasinski2019demonstration}, marking significant milestones in the fields of quantum computation and information \cite{zhou2000Methodology,son2001conclusive,Lee2021Quantum,chitambar2023duality}.
Building upon these foundational achievements, quantum teleportation has accelerated the development of advanced quantum technologies, including entanglement swapping \cite{pan1998experimental}, quantum repeaters essential for long-distance quantum communication \cite{briegel1998quantum,lee2019fundamental}, quantum gate teleportation \cite{gottesman1999demonstarting}, quantum cryptography \cite{pirandola2020advances}, and measurement-based computation \cite{raussendorf2001one}.
Such innovations underscore the protocol's critical role in enabling secure and efficient quantum networks \cite{liao2018satellite,daiss2021quantum,pompili2021realization}, paving the way for the realization of a future quantum internet \cite{kimble2008quantum,wehner2018quantum}.

However, implementation of quantum communication encounters various challenges in the real world, leading to the proposal of different modified teleportation protocols to overcome these impediments.
For example, catalytic quantum teleportation was proposed to overcome inevitable practical noise in resource states, utilizing entanglement states that are not consumed or degraded during the process \cite{jonathan1999entanglement}.
Lipka-Bartosik \textit{et al.} \cite{lipka2021catalytic} proved that this protocol could achieve teleportation fidelity equal to that of noiseless teleportation.
Moreover, based on linear optics, the maximum probability of successfully distinguishing Bell states is limited to 50\%, significantly reducing the efficiency of teleportation \cite{calsamiglia2001maximum}.
Through the use of ancillary photons, the teleportation scheme proposed by Knill, Laflamme, and Milburn (KLM) enables asymptotically perfect state transmission\cite{knill2001scheme}, while other experiment recently has demonstrated Bell-state measurements exceeding this limit \cite{bayerbach2023bell}.

An asymptotic teleportation scheme that allows the receiver to make selections without performing quantum corrections has been proposed, known as port-based teleportation (PBT) \cite{ishizaka2008asymptotic}.
This approach provides a universal programmable processor in a simple and natural manner.
With its application potential in cryptography \cite{Beigi2011simplified}, holography \cite{may2022complexity}, quantum computing \cite{sedlak2019optimal,quintino2019reversing}, and PBT contributes significantly to quantum communication.
It also sheds light on the non-local measurements of multipartite states and advances understanding of communication complexity \cite{buhrman2016quantum} and quantum channels \cite{Pirandola2019fundamental}. 
Current research efforts are focused on modifying and optimizing PBT \cite{Mozrzymas2018optimal,jeong2020generalization,studzinski2022efficient,Strelchuk2023minimal}, expressing it as efficient quantum circuits \cite{grinko2023efficient}, and analysing its performance against noise \cite{kim2024port}.

Accordingly, various quantum teleportation schemes have been proposed and developed with the aim of surmounting experimental constraints or targeting specific applications.
Such individuality and diversity prompt us to ask the ensuing question:
`Is it feasible to transition between different teleportation schemes through incremental adjustments of their parameters?'
Exploring this possibility could unveil underlying connections between seemingly disparate teleportation mechanisms, offering a unified perspective on quantum communication.
Our investigation seeks to not only validate the theoretical feasibility of such transitions but also to understand potential enhancements to teleportation efficiency and flexibility.

In this paper, we take the first steps by starting with an analysis on the effects of altering joint positive operator-valued measure (POVM) elements.
Given the rigorously defined mathematical framework of the PBT protocol and its comprehensive analysis in prior research, we delve into studying a novel form of asymptotic teleportation inspired by it, which we term port-based quantum correction teleportation (PBQCT).
This approach demands the receiver to perform quantum corrections following port selection to complete the process.
Moreover, PBQCT scheme encompasses PBT protocol when the number of POVM elements is at its minimum, and transitions to parallel ST protocol as the number of POVM elements increases and reaches the maximum.

In particular, within two-dimensional system, the PBQCT scheme is categorized into four groups with identical performance. We provide proof of this classification and analytically examine the protocols within all groups.
Among these groups, one group's protocol features POVM elements that can be simply expressed in the computational basis and possess a form amenable to the application of stabilizer formalism.
In higher dimensions, we identify protocols that possess properties similar to those of the group protocol, and we analyze these to assess their scalability and applicability across different dimensional systems.

This paper is organized as follows.
In section \ref{sec:teleportation}, we introduce and clarify the teleportation scenario we are considering.
At the end of section \ref{subsec:model}, we describe ST and PBT from the perspective of our scenario.
Furthermore, we specify our scenario and provide an asymptotically perfect PBQCT scheme in section \ref{subsec:pbqct}.
In section \ref{sec:pbqct_2}, we analyze PBQCT within a qubit system.
We show that PBQCT protocols can be categorized into four distinct groups based on their identical performance.
Furthermore, we find the analytic form of the joint measurement and assess its performance.
In section \ref{sec:high}, we extend our analysis to a qudit system.
We prove that every protocol in PBQCT scheme achieves perfect transmission in the asymptotic limit of a large number of ports in section \ref{subsec:asymptotic}.
Moreover, we generalize and examine one of the group's protocols from a qubit system to higher dimensions in section \ref{subsec:gen_pbqct_2}.

\section{
    \label{sec:teleportation}
    Port-based Quantum Correction Teleportation: The Scenario and Protocol
    }

This section begins by delineating the teleportation scenario being examined in our study.
The ST and PBT protocols serve as specific examples within this scenario, and we revisit and describe these protocols from our perspective.
Lastly, we provide a brief overview of novel deterministic and asymptotically perfect teleportation scheme inspired by PBT. The details and main results will be covered in sections \ref{sec:qubit} and \ref{sec:high}.

\subsection{
    \label{subsec:model}
    Teleportation Scenario
    }

    \begin{figure*}%[h]
	\centering
        \includegraphics[width=\linewidth]{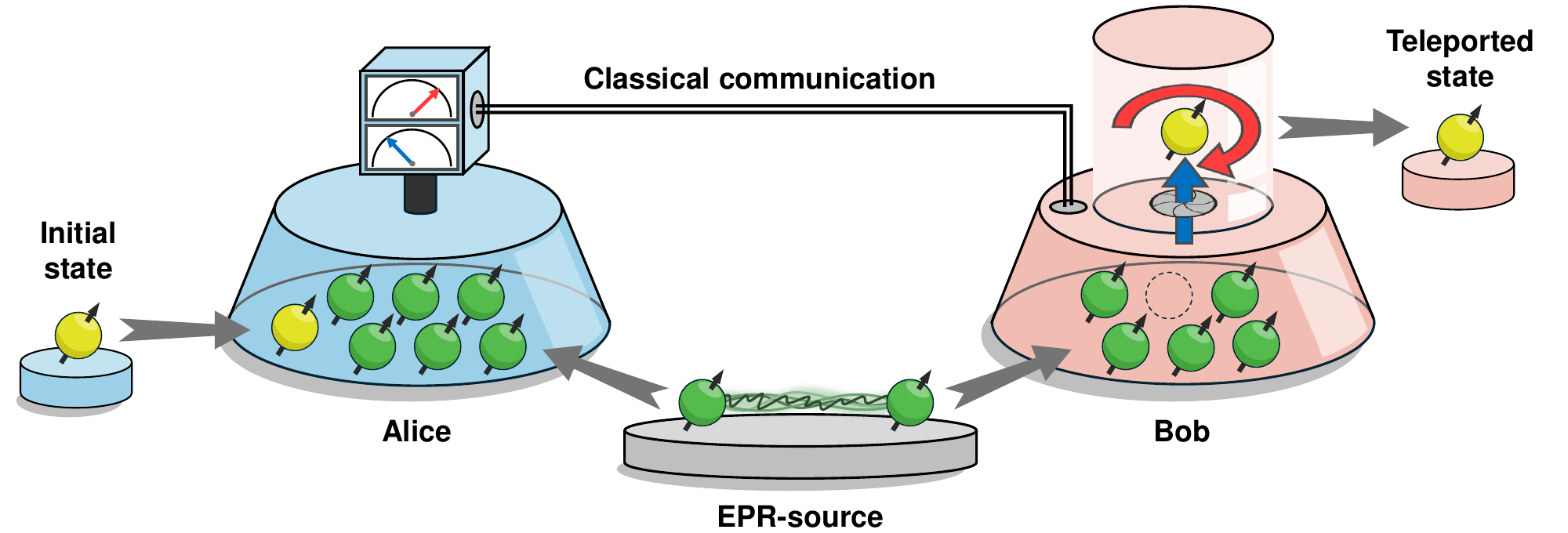}
        \caption{\label{fig:tel}
        The scheme for port-based quantum correction teleportation (PBQCT).
        Alice wishes to teleport a quantum state, encoded in her yellow qudit, to Bob.
        They share $N$ copies of maximally entangled pair of particles, shown as green qudits, originating from an EPR source.
        Alice then conducts a joint POVM, inspired from PBT protocol, on her yellow qudit and her bundles of green qudits.
        The measurement leads to two types of outcomes that convey classical and quantum correction information.
        She gets two type of outcomes, indicating classical and quantum correction information.
        Upon receiving these outcomes via classical communication, Bob selects the corresponding qudit and applies a quantum operation to reconstruct Alice's initial state.
        In the asymptotic limit of large $N$, the initial state is perfectly teleported.}
    \end{figure*}

The teleportation scheme under consideration through our study is depicted by Fig. \ref{fig:tel}.
Consider Alice and Bob, who are spatially separated from each other.
Each of them receives one part of a maximally entangled qudit pair, emitted by a $D$-dimensional Einstein-Podolsky-Rosen (EPR) source. The state of this qudits pair can be represented as 
    \begin{equation}
    \label{eq:def_max_ent}
        \Ket{\Psi^{(0,0)}}
        =\frac{1}{\sqrt{D}}\sum_{j=0}^{D-1}
        \ket{j}\otimes\ket{j}.
    \end{equation}
This preparation process is repeated $N$ times. 
During each iteration, Alice and Bob store each pair of qudits as $A_i$ and $B_i$, where $i=1,2,...,N$.
We refer to $A_i$ and $B_i$ as the $i$th port of Alice and Bob’s resource systems, given by $\vec{A} := \{A_1, A_2, ..., A_N \}$ and $\vec{B} := \{B_1, B_2, ..., B_N \}$, respectively.
Furthermore, Alice prepares a $D$ dimensional qudit $A_0$ that holds an unknown quantum state $\ket{\rho}$.

Subsequently, Alice needs to conduct a suitable joint POVM measurement on $A_0$ and $\vec{A}$ to transmit the unknown state $\ket{\rho}$ to Bob.
We consider the square-root measurement (SRM) as a viable choice for this joint measurement.
The SRM has been shown to provide good, but not optimal, performance in discriminating among a given set of states \cite{hausladen1994pretty,leditzky2022optimality}.
Furthermore, SRM plays a crucial role in the modified teleportation protocol, known as PBT, which eliminates the need for quantum correction.
Details of PBT are deferred to the latter part of this section.
The elements of SRM are defined as
    \begin{equation}
    \label{def:gen_srm}
        \Sigma^{m}:=
        G^{-\frac{1}{2}}
        g^{(m)}
        G^{-\frac{1}{2}}\;\;;\;\;
        G:=\sum_{m\in\mathfrak{m}}g^{(m)},
    \end{equation}
where $g^{(m)}$ denotes a signal state \cite{peres1991optimal} and denote $G$ as signal sum operator.
In Eq. (\ref{def:gen_srm}), $\mathfrak{m}$ represents a set of measurement outcomes and consists of a total of $|\mathfrak{m}|$ elements. Without loss of generality, we suppose the classical outcomes to be an integer or a list of integers. 

Since the rank of $G$ cannot be always full, the inversion is defined on the support of $G$.
Consequently, SRM fails to satisfy the conditions required for a POVM, specifically that the sum of its elements should equal the identity operator.
To construct a joint POVM measurement based on SRM, a projection operator corresponding to $G$'s null space is added to each signal state.
This operator is scaled by the inverse of the signal set size, ensuring that the adjusted elements contribute appropriately to the POVM construction.
Through mathematical representation, the POVM elements are characterized as
    \begin{equation}
    \label{def:gen_POVM}
        \Pi^m
        =
        \Sigma^m
        +\Delta
        \;;\;
        \Delta
        :=
        \frac{1}{|\mathfrak{m}|}
        \left(
        I
        -
        \sum_{m\in\mathfrak{m}}
        \Sigma^m
        \right),
    \end{equation}
where the summation of all elements equals the identity operator $I$, resulting in $\sum_{m\in\mathfrak{m}}\Pi^m=I$.
Significantly, the presence of the additional term $\Delta$ has no effect on the entanglement fidelity, as evidenced by $\mathrm{Tr}\Bigl[g^{(m)}\Delta\Bigr]=0$.

After the joint measurement, Alice sends her outcome $m\in\mathfrak{m}$ to Bob over a classical channel.
Upon receiving the classical message $m$, Bob initiates the decoding process. 
He performs a decoding operation on his resource qudits $\vec{B}$, utilizing the channel $\mathcal{D}^{m}$, in an attempt to recover the unknown state $\ket{\rho}$ in one of his qudits.
The decoding channel $\mathcal{D}^{m}$ generally operates through a mixture of classical means and quantum mechanisms. In certain teleportation protocols, however, only one approach may be required, or the two processes may be applied sequentially.
The teleportation channel $\Lambda(\rho)$
can be mathematically represented as
    \begin{widetext}
    \begin{equation}
        \Lambda(\rho)=\sum_{m\in{\mathfrak{m}}}\bigl(
        \Tr_{A_0\vec{A}}\otimes\mathcal{D}^{m}_{\vec{B}}
        \bigr)
        \left[
        \left(
        \sqrt{ \Pi^{m}}_{A_0\vec{A}}
            \otimes I_{\vec{B}}
        \right)
        \left(
        \rho_{A_0}
            \otimes \Psi_{\vec{A}\vec{B}}
        \right)
        \left(
        \sqrt{ \Pi^{m}}^\dagger_{A_0\vec{A}}
            \otimes I_{\vec{B}}
        \right)
        \right].
        \label{def:gen_tel_ch}
    \end{equation}
    \end{widetext}
The resource states $\ket{\Psi}_{\vec{A}\vec{B}}$ utilized in the Eq. (\ref{def:gen_tel_ch}) are formulated as
    \begin{equation}
    \label{def:gen_resource_st}
        \frac{1}{\sqrt{D^N}}
        \sum_{j_1=0}^{D-1}...\sum_{j_N=0}^{D-1}
        \left[
        \left(\bigotimes_{n=1}^{N}\ket{j_n}_{A_n}\right)
        \left(\bigotimes_{n'=1}^{N}\ket{j_{n'}}_{B_{n'}}\right)
        \right],
    \end{equation}
depicting the $\Ket{\Psi^{(0,0)}}^{\otimes N}$ state, but with the sequences of $\vec{A}$ and $\vec{B}$ rearranged for clarity.
If the context clearly prevents any confusion, the spaces in which states, operators, and channels act may be denoted in italics with subscripts corresponding to their symbols without further specification.
Furthermore, we define a state represented by a density operator and use the same symbol for both the density operator and its corresponding pure state. This notation simplifies our discussion and analysis, as exemplified by the resource states $ \Psi :=\ket{\Psi}\bra{\Psi}$ in Eq. (\ref{def:gen_tel_ch}).

The efficiency of a teleportation channel, $\Lambda$, is determined by its teleportation fidelity. This fidelity is calculated by integrating over the uniform distribution $d\rho$ of all pure states in a $d$-dimensional space:
    \begin{equation}
    \label{eq:def_tel_fid}
        f(\Lambda):=\int\bra{\rho}\Lambda
            \bigl(\rho\bigr)
            \ket{\rho}
            d\rho,
    \end{equation}
where the fidelity ranges from $0$ to $1$.
Teleportation fidelities that exceed the classical limit of $2/(D+1)$ demonstrate quantum superiority \cite{badziag2000local}.
Further analysis includes entanglement fidelity, which assesses quantum channel noise and entanglement preservation, defined by:
    \begin{equation}
    \label{eq:def_ent_fid}
        F(\Lambda):=
        \bra{\Psi^{(0,0)}}
        \Bigl[
        (\Lambda\otimes\mathbf{1})\ket{\Psi^{(0,0)}}\bra{\Psi^{(0,0)}}
        \Bigr]
        \ket{\Psi^{(0,0)}},
    \end{equation} 
with $\mathbf{1}$ representing the identity channel.
A fundamental relationship between teleportation and entanglement fidelity, as identified by Horodecki \textit{et al.} \cite{horodecki1999general}, follows the equation:
    \begin{equation}
    \label{eq:rel_tel_fid_and_ent_fid}
        f(\Lambda)=\frac{F(\Lambda)D+1}{D+1}.
    \end{equation}

The teleportation scenario considered in this study encompasses both the initially proposed ST and PBT.
In the remainder of this section, we describe how both protocols are adapted to our specific scenario.
To ensure the clarity, we adopt a consistent notation for operators and sets associated with each protocol, using roman subscripts. For example, the outcome set for the ST protocol is denoted as $\mathfrak{m}_{\mathrm{ST}}$, and a POVM element for the PBT protocol is represented as $ \Pi^m_{\mathrm{PBT}}$.

\textit{Standard Teleportation}---
As first proposed by Bennett \textit{et al.} in 1993 \cite{bennett1993teleporting}, ST protocol transmit an unknown quantum state through dual classical and EPR channels.
To adapt this protocol for systems in $D$ dimensions, we begin by introducing a generalization of the Pauli operators, known as the Weyl-Heisenberg operators.
Theses operators $ W^{p,q}$ are defined as
    \begin{equation}
         W^{(p,q)}:= P^p Q^q\;;\;p,q=0,1,2,...,D-1\;,
    \end{equation}
where the shift operator $ P $ and the clock operator $ Q $ are given respectively by
    \begin{equation}
        \label{def:shft_mat}
         P :=\sum_{j=0}^{D-1}\ket{j}\bra{j\oplus 1},
    \end{equation}
and         
    \begin{equation}
         Q :=\sum_{j=0}^{D-1}e^{\frac{2\pi i}{D}j}\ket{j}\bra{j}.
    \end{equation}
Remarkably, while the Weyl-Heisenberg operators preserve the unitarity and tracelessness as Pauli operators do, Hermiticity is not maintained in dimensions higher than two.

The ST protocol consumes a single pair of EPR qudits ($N=1$), utilizing generalized Bell measurement as joint POVM measurement. The measurement elements are expressed as     
    \begin{equation}
    \label{def:ST_POVM}
        \Pi^{(p,q)}_{\mathrm{ST}}:= \Psi^{(p,q)},
    \end{equation}
corresponding to generalized Bell states, defined as
    \begin{eqnarray}
        \Ket{\Psi^{(p,q)}}
        &=&\sum_{j=0}^{D-1}e^{\frac{2\pi i}{N}jp}
        \ket{j}\otimes\ket{j\oplus q}
        \nonumber
        \\
        &=&\bigl( W^{(p,q)}\otimes I \bigr)\Ket{\Psi^{(0,0)}}.
        \label{eq:ST_POVM_sts}
    \end{eqnarray}
The set of outcomes, $\mathfrak{m}_{\mathrm{ST}}$, is defined as
    \begin{equation}
        \mathfrak{m}_{\mathrm{ST}}:=\{(p,q)\;:\;p,q=0,1,2,...,D-1\},
    \end{equation}
Notably, the signal sum operator $ G $ in Eq. (\ref{def:gen_srm}) simplifies to an identity operator, resulting the POVM elements in Eq. (\ref{def:gen_POVM}) to be directly expressed as shown in Eq. (\ref{def:ST_POVM}).
These elements can be regarded as projection operators onto an EPR state rotated with $ W^{(p,q)}$ in a single qudit, as expressed in Eq. (\ref{eq:ST_POVM_sts}). Consequently, the decoding operation can be succinctly described as a quantum correction, denoted by
\begin{equation}
\mathcal{D}^{p,q}_{\mathrm{ST}}(X) :=  W^{(p,q)}X{W^{(p,q)}}^\dagger.
\end{equation}
It is essential to note that the ST protocol simplifies the process by requiring only quantum corrections, with no further classical actions necessary.
Since ST protocol guarantees a perfect state transfer, both entanglement and teleportation fidelities are given as 
    \begin{equation}
        F(\Lambda_{\mathrm{ST}})=
        f(\Lambda_{\mathrm{ST}})=1,
    \nonumber
    \end{equation}
where $\Lambda_{\mathrm{ST}}$ denotes the teleportation channel of ST.

\textit{Port-based Teleportation}---
Ishizaka and Hiroshima \cite{ishizaka2008asymptotic} proposed a modified teleportation protocol in which Bob is not required to perform any quantum operations.
To achieve this advantage, PBT employs $N$ pairs of EPR states, leveraging a more complex joint measurement than the Bell measurement used in ST. 
The POVM elements, $ \Pi^{m}_{\mathrm{PBT}}$, are described by exploiting Eq. (\ref{def:gen_POVM}) with signal state given as
    \begin{eqnarray}
    \label{def:PBT_sgn_sts}
        \left[ g^{(i)}_{\mathrm{PBT}}
        \right]_{A_0\vec{A}}
        &:=&
        \Tr_{\bar{B}_i}
        \Bigl[
             \Psi_{\vec{B}\vec{A}}
        \Bigr]
        \Biggr|_{B_i\rightarrow A_0}
        \nonumber\\
        &=&\frac{1}{D^{N-1}} \Psi^{(0,0)}_{A_0A_i}\otimes I_{\bar{A}_i},
    \end{eqnarray}
where $ I_{\bar{A}_i}$ represents the identity operator acting on the remaining qubits $\bar{A}_i:=\vec{A}\backslash\{A_i\}$.
It is important to note that the signal state $ g^{(i)}_{\mathrm{PBT}}$  is obtained by partially tracing out all qubits except the $n$th qubit from the resource state, as defined in Eq. (\ref{def:gen_resource_st}).
The set of outcomes, $\mathfrak{m}_{\mathrm{PBT}}$, is defined as 
    \begin{equation}
        \mathfrak{m}_{\mathrm{PBT}}
        :=
        \left\{
            i\;:\;i=1,2,...,N
        \right\}.
    \end{equation}
In this protocol, Bob requires no quantum corrections; instead, he simply selects a qubit from $\vec{B}$ based on the measurement outcome. The classical decoding channel $\mathcal{D}^i_{\mathrm{PBT}}:\vec{B}\rightarrow B_i$ is then defined by:
    \begin{equation}
        \mathcal{D}^i_{\mathrm{PBT}}(X_{\vec{B}})
        :=
        \mathrm{Tr}_{\bar{B}_i}
        \left[
        X_{\vec{B}}
        \right],
    \end{equation}
where $\mathrm{Tr}_{\bar{B}_i}$ signifies the partial trace over the complement qubit space $\bar{B}_i$.

The PBT protocol can perfectly transmit an unknown state as the number of ports $N$ approaches infinity.
However, a finite number of ports leads to information loss and reduces teleportation fidelity.
Studzi\'{n}ski \textit{et al.} \cite{studzinski2017port} determined the exact form of entanglement fidelity for $D$-dimensional $N$-PBT.
Specifically, for qubit systems ($D=2$), the entanglement fidelity is given by \cite{ishizaka2008asymptotic,ishizaka2009quantum}
    \begin{equation}
    \label{eq:ent_fid_PBT}
        F(\Lambda_{1}
        )=\frac{1}{2^{N+3}}\sum^N_{k=0}
        \left(
        \frac{N-2k-1}{\sqrt{k+1}}+\frac{N-2k+1}{\sqrt{N-k+1}}
        \right)^2
        {N \choose k},
    \end{equation}
with $\Lambda_{1}$ representing the PBT teleportation channel.
In the asymptotic limit of $N\rightarrow \infty$, entanglement fidelity approaches
    \begin{equation}
        F(\Lambda_{1})\rightarrow1-\frac{3}{4N},
    \nonumber
    \end{equation}
while teleportation fidelity approaches
    \begin{equation}
        f(\Lambda_{1})\rightarrow1-\frac{1}{2N}
    \nonumber
    \end{equation}
according to Eq. (\ref{eq:rel_tel_fid_and_ent_fid}).

\subsection{
    \label{subsec:pbqct}
    Port-based Quantum Correction Teleportation
    }

    \begin{figure}[h]
	\centering
        \includegraphics[width=\linewidth]{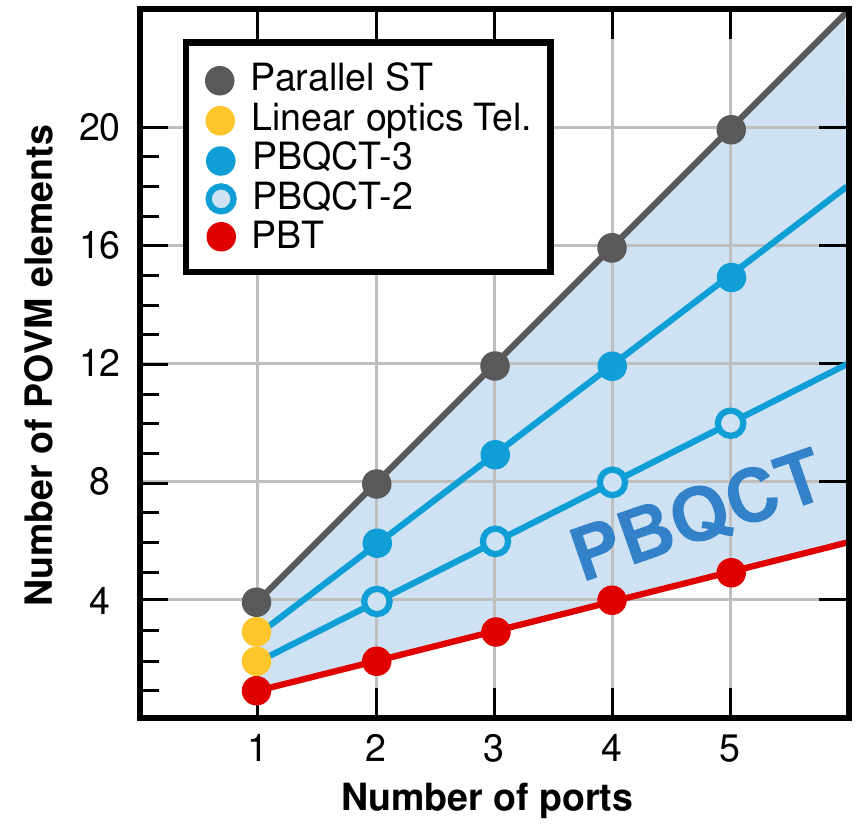}
        \caption{\label{fig:tel_diag}
        Asymptotic teleportation scheme diagram for qubit systems with respect to the number of ports and POVM elements.
        The blue shaded region depicts the PBQCT scheme.
        Protocols with the minimum number of POVM elements, including port-based teleportation (PBT) protocol, are denoted by red dots.
        The upper boundary of the domain, corresponding to the parallel standard teleportation (ST) protocol, is indicated with gray dots.
        For a single port, teleportation via linear optics is interpreted as two distinct PBQCT protocols, shown as yellow dots, depending on the approach to POVM.
        Incrementing the number of ports extends the protocols into  PBQCT-2 and PBQCT-3 protocols, indicated by open and closed blue dots, respectively.}
    \end{figure}

Drawing upon the insights from the PBT protocol, we introduce an asymptotic teleportation scheme rooted in our scenario.
To understand this scheme, we revisit the PBT protocol, which is characterized by its requirement for only classical corrections for port-selection without the need for any quantum operations.
This distinctive feature of PBT stems from its signal state configuration, as defined in Eq. (\ref{def:PBT_sgn_sts}).
In the asymptotic limit as $N$ approaches infinity, the operators $ G $ and $ \Delta $, crucial for distorting and completing the POVM measurement, converge to the identity and zero operators, respectively.
This convergence directly transforms the joint POVM elements into the signal states.
Furthermore, it becomes unnecessary to consider any qubits other than $A_0$, $A_n$, and $B_n$ upon observing the outcome $n$. Thus, the state of $A_0$ and $A_n$ qubits collapse to a Bell state, identical to the resource state between $A_n$ and $B_n$ qubits, as defined in Eq. (\ref{eq:def_max_ent}).
This implies that once a port is selected, state transmission is ensured solely through a projection onto a prepared Bell state, rendering Bob's operation essentially an identity operation.
Building upon this foundation, we extend the signal set used in PBT to include teleportation protocols that necessitate not only port-selection but also additional quantum corrections.
We refer to this generalized measurement approach as port-based quantum correction teleportation, abbreviated as PBQCT.

Originating from PBT, our scheme inherits the characteristic of achieving perfect fidelity in the asymptotic limit which we will prove in section \ref{subsec:asymptotic}.
Furthermore, given that PBQCT protocols utilize the expanded signal set of PBT, they can be systematically organized and interconnected based on the number of POVM elements.
Specifically, for qubit systems, Fig. \ref{fig:tel_diag} demonstrates diagram of asymptotic teleportation scheme transitions with changes in the number of ports and POVM elements.
The area highlighted in blue represents the domain where PBQCT protocols are applicable.
The range begins with the minimum number of POVM elements, where the count matches the number of ports and includes PBT, plotted by red dots.
It extends to the maximum, with the count reaching the square of the dimension.
This transition evolves into parallel ST functioning as independent protocols across ports as detailed in section \ref{sec:qubit}.
Notably, this transition is represented by gray dots.
Therefore, the PBQCT scheme ensures the preservation of asymptotic perfect fidelity, permitting a fluid transition between both protocols by gradually modifying the signal set.
%그림의 파란색이 PBQCT임을 명시.

In Fig. \ref{fig:tel_diag}, the two yellow dots signify the linear optics teleportation protocol that employs a single EPR pair.
Due to constraints in present Bell-state measurement techniques within linear optics, only two of the four Bell states are identifiable.
Owing to these constraints, the teleportation protocol can be interpreted in two different ways, as deterministic protocols with two and three POVM elements.
Considering the two indistinguishable Bell states as a single independent element of a POVM, the protocol operates through a three-element POVM. Conversely, if the space projected by these two states is regarded as $\Delta$ in Eq. (\ref{def:gen_POVM}), it results in a protocol utilizing a two-element POVM.
Each protocol can be extended into PBQCT-2 and PBQCT-3 by doubling and tripling the number of ports, respectively.
These protocols are indicated by blue open and closed dots in Fig. \ref{fig:tel_diag}.
We detail these protocols further in Sections 3 and 4.

In our initial exploration of PBQCT, we investigate signal sets that remain invariant under the action of any permutation on the qubits $\vec{A}$, following the approach in PBT.
This implies that every qubit is treated uniformly, without any particular distinction.
Furthermore, we impose the condition that all signal states become one of the generalized Bell states when the identity part is traced out.
This constraint is motivated by the fact that SRM is recognized as a highly effective measurement, approaching optimality, particularly for signal sets with orthogonal states.
Moreover, this constraint enables the scheme to achieve asymptotic perfect teleportation fidelity.
More detailed investigations and results are provided in sections \ref{sec:qubit} and \ref{sec:high} for two-dimensional and higher dimensional systems, respectively.

\section{
    \label{sec:qubit}
    Qubit systems
    }

The correction process required by Bob after receiving the outcomes is determined by the signal set of the SRM, and each protocols are considered distinct.
Specifically in qubit systems, we demonstrate how PBQCTs with varying signal sets can be classified by entanglement fidelity from the properties of the Pauli group.
To begin with, we assume that every PBQCT can have signal states solely in the form of
    \begin{eqnarray}
         g^{(i,s)}:=
        \frac{1}{2^{N-1}} \Psi^s_{A_0A_i}\otimes I_{\bar{A}_i}
        ,
    \end{eqnarray}
where $ \Psi^s:=\ket{\Psi^s}\bra{\Psi^s}$ denotes Bell states, which are
    \begin{eqnarray}
        \ket{\Psi^0}&:=&
        \frac{\ket{0}\otimes\ket{0}+\ket{1}\otimes\ket{1}}
        {\sqrt{2}},
        \nonumber
        \\
        \ket{\Psi^1}&:=&
        \frac{\ket{0}\otimes\ket{1}+\ket{1}\otimes\ket{0}}
        {\sqrt{2}},
        \nonumber
        \\
        \ket{\Psi^2}&:=&
        \frac{\ket{0}\otimes\ket{1}-\ket{1}\otimes\ket{0}}
        {\sqrt{2}i},
        \nonumber
    \end{eqnarray}
and
    \begin{eqnarray}
        \ket{\Psi^3}&:=&
        \frac{\ket{0}\otimes\ket{0}-\ket{1}\otimes\ket{1}}
        {\sqrt{2}}.
        \nonumber
    \end{eqnarray}
Bell states can be transformed into one another using local Pauli operators, allowing them to be expressed as
    \begin{eqnarray}
        \ket{\Psi^s}
        =\left(
         \sigma_s\otimes \sigma_0
        \right)\ket{\Psi^0}.
    \label{eq:qb_pl_bell}
    \end{eqnarray}
    
Here, we assume that the signal set of all PBQCT protocols we consider is closed under the action of any permutation on the qubits $\vec{A}$;
then we can represent the outcome sets of a PBQCT as
    \begin{eqnarray}
    \label{def:qb_sgset}
        \mathfrak{m}:=\left\{
            (i,s)
            :
            i\in\{1,2,...,N\},\;
            s\in\mathfrak{s}
        \right\},
    \end{eqnarray}
where $\mathfrak{s}$ is a nonempty subset of Bell states' numbers $\{0,1,2,3\}$.
By applying the definition of SRM elements (\ref{def:gen_srm}) and POVM elements, we can obtain the SRMs' and POVMs' elements of PBQCT.
Furthermore, by exploiting Eq. (\ref{eq:def_ent_fid}), we can obtain the entanglement fidelity for the channel $\Lambda_\mathfrak{s}$ of PBQCT with outcome set $\mathfrak{s}$ as 
    \begin{subequations}
    \begin{eqnarray}
        F(\Lambda_\mathfrak{s})
        &=&
        \mathrm{Tr}
        \left[
             \Psi^0\left(\Lambda_\mathfrak{s}
                \otimes \mathbf{1}_D\right)
             \Psi^0
        \right]
        \nonumber
        \\
        &=&
        \sum_{i=1}^{N}
        \sum_{s\in\mathfrak{s}}
        \mathrm{Tr}
        \Bigl[
        \left(
         \Psi^0_{B_iD}
        \otimes \Pi^{(i,s)}_{A_0\vec{A}}
        \right)
        \nonumber
        \\
        & &\qquad\qquad\quad
            \cdot\left( g^{(i,s)}_{B\vec{A}}
            \otimes \Psi^0_{A_0D}
        \right)
        \Bigr]
        \nonumber
        \\
        &=&
        \frac{1}{2^2}
        \sum_{i=1}^{N}
        \sum_{s\in\mathfrak{s}}
        \mathrm{Tr}
        \left[
         \Pi^{(i,s)}_{B\vec{A}}
         g^{(i,s)}_{B\vec{A}}
        \right]
        \label{eq:qb_ent_fid_3}
        \\
        &=&
        \frac{1}{2^2}
        \sum_{i=1}^{N}
        \sum_{s\in\mathfrak{s}}
        \mathrm{Tr}
        \left[
         G^{-\frac{1}{2}}
         g^{(i,s)}
         G^{-\frac{1}{2}}
         g^{(i,s)}
        \right].
        \label{eq:qb_ent_fid_4}
    \end{eqnarray}
    \end{subequations}
The acting space of $ \Pi $ is changed to the Hilbert space $\mathcal{H}_B\otimes\mathcal{H}_{\vec{A}}$ in Eq. (\ref{eq:qb_ent_fid_3})
using the twirling invariance of Bell state, as expressed by
    \begin{eqnarray}
        \ket{\Psi^0}=
        \left(
         u \otimes u^T
        \right)
        \ket{\Psi^0},
    \label{eq:qb_twl}
    \end{eqnarray}
with $ u $ representing an arbitrary unitary operator.
Since all operators act within the same space, we omitted the notation of operating space in Eq. (\ref{eq:qb_ent_fid_4}).

We move on to demonstrating  characteristics of PBQCT scheme in qubit systems.
These properties result from the fact that there always exists a unitary operator that can transform a signal set into another signal set of the same size, and it is invariant under any permutation of the qubits $\vec{A}$.
To find these unitary operators, we convert the signal state by
    \begin{subequations}
    \begin{eqnarray}
         g^{(i,s)}_{A_0\vec{A}}
        &=&
        \left(
         \sigma^s_{A_0}\otimes  I_{\vec{A}}
        \right)
         g^{(i,0)}_{A_0\vec{A}}
        \left(
         \sigma^s_{A_0}\otimes  I_{\vec{A}}
        \right)
        \label{eq:sym_sgl_1}
        \\
        &=&
         u^\dagger_{A_0\vec{A}}
        \left(
         u_{A_0} \sigma^s_{A_0}
         u^\dagger_{A_0}\otimes I_{\vec{A}}
        \right)
        \nonumber
        \\
        & &\cdot
         g^{(i,0)}_{A_0\vec{A}}
        \left(
         u_{A_0} \sigma^s_{A_0}
         u^\dagger_{A_0}\otimes I_{\vec{A}} 
        \right)
         u_{A_0\vec{A}},
        \label{eq:sym_sgl_2}
    \end{eqnarray}
    \end{subequations}
where $ u $ denotes an arbitrary unitary operator, and $ u_{A_0\vec{A}}= u_{A_0}\otimes( u^{T})^{\otimes\vec{A}}$.
The Pauli operators in Eq. (\ref{eq:sym_sgl_1}) are derived from the relationship involving Bell states as shown in Eq. (\ref{eq:qb_pl_bell}).
Equation (\ref{eq:sym_sgl_2}) is obtained by applying $ u_B u_B^\dagger$, which is identity, between operators and using twirling invariance (\ref{eq:qb_twl}).
By the same method used to omit $ u $ and $ u^\dagger$ on the left and right of signal state, respectively, we can introduce another arbitrary unitary operator $ v $, and express $ g^{(i,s)}_{A_0\vec{A}}$ as follows:
    \begin{eqnarray}
         g^{(i,s)}_{A_0\vec{A}}&=& W^\dagger_{A_0\vec{A}}
        \left(
         u_{A_i} \sigma^s_{A_i}
         u^\dagger_{A_i} v^\dagger_{A_i}\otimes  I_{A_0\bar{A}}
        \right)
        \nonumber
        \\
        & &\cdot
         G^i_{A_0\vec{A}}
        \left(
         v_{A_i} u_{A_i}
         \sigma^s_{A_i} u^\dagger_{A_i}
        \otimes  I_{A_0\bar{A}}
        \right)
         W_{A_0\vec{A}},
        \label{eq:qb_sgl_trns_W}
    \end{eqnarray}
where $ W_{A_0\vec{A}}= u_{A_0}\otimes( u^{T} v^{T})^{\otimes\vec{A}}$.

In the context of the signal state as described in Eq. (\ref{eq:qb_sgl_trns_W}), we focus on the transformation of the Pauli operator due to $ u $ and $ v $. We denote this transformation as $T$, defined by
    \begin{eqnarray}
    \label{eq:pbqct_qb_pauli_trans}
        T
        \left(
         \sigma^s
        \right)
        :=
         u  \sigma^s u^\dagger v^\dagger.
    \end{eqnarray}
By constraining $ u $ to be a Clifford operator and $ v $ to be a Pauli operator in the transformation $T$, every one-to-one corresponding function can be generated, with both the domain and codomain as the set of Pauli operators $\{ \sigma^0, \sigma^1, \sigma^2, \sigma^3\}$.
Note that the overall phase difference is neglected, as the transformation is applied once and conjugated in Eq. (\ref{eq:qb_sgl_trns_W}).
Finally, Eq. (\ref{eq:qb_sgl_trns_W}) can be rewritten as
    \begin{eqnarray}
         W_{A_0\vec{A}}
         g^{(i,s)}_{A_0\vec{A}}
         W^\dagger_{A_0\vec{A}}
        &=&
        \left(
         \sigma^{s'}_{A_0}\otimes I_{\vec{A}}
        \right)
         g^{(i,0)}_{A_0\vec{A}}
        \left(
         \sigma^{s'}_{A_0}\otimes I_{\vec{A}}
        \right)
        \nonumber
        \\
        &=&
         g^{(i,s')}_{A_0\vec{A}},
        \label{eq:qb_sgl_trns_last}
        \end{eqnarray}
assuming $T$ transforms the Pauli operator number $s$ to $s'$.
Thus, Eq. (\ref{eq:qb_sgl_trns_last}) demonstrates that a unitary operator exists which transforms the signal set into another, and this transformation is invariant under permutation among the qubits $\vec{A}$.

The existence of unitary operators ensure that all entanglement fidelity of measurements within a chosen PBQCT are identical.
Consider the outcomes $(i,s)$ and $(i,s')$ within this PBQCT.
The entanglement fidelity for a channel with measurement $(i,s)$ can be equivalently transformed to that with $(i,s')$, as demonstrated below:
    \begin{eqnarray}
        F(\Lambda_{i,s})
        &=&
        \frac{1}{2^2}\mathrm{Tr}\left[
         G^{-\frac{1}{2}}
         g^{(i,s)}
         G^{-\frac{1}{2}}
         g^{(i,s)}
        \right]
        \nonumber
        \\
        &=&
        \frac{1}{2^2}\mathrm{Tr}\Bigl[
        \left(
         W 
         G^{-\frac{1}{2}}
         W^\dagger
        \right)
        \left(
         W 
         g^{(i,s)}
         W^\dagger
        \right)
        \nonumber
        \\
        & &\qquad\quad\cdot
        \left(
         W 
         G^{-\frac{1}{2}}
         W^\dagger
        \right)
        \left(
         W 
         g^{(i,s)}
         W^\dagger
        \right)
        \Bigr]
        \nonumber
        \\
        &=&
        \frac{1}{2^2}\mathrm{Tr}\left[
         G^{-\frac{1}{2}}
         g^{(i,s')}
         G^{-\frac{1}{2}}
         g^{(i,s')}
        \right]
        \nonumber
        \\
        &=&
        F(\Lambda_{i,s'}),
        \label{eq:fid_meas_ch}
    \end{eqnarray}
where $ W $ is chosen specifically to transform $ g^{(i,s)}$ into $ g^{(i,s')}$, thus ensuring the preservation of the signal set.
This principle applies equally across PBQCT protocols with signal sets of the same size. For different but equally sized sets of Bell outcomes $\mathfrak{s}$ and $\mathfrak{s}'$, the entanglement fidelity of the channel $\Lambda_{\mathfrak{s}}$ over the set of Bell outcome $\mathfrak{s}$ can be transformed with same manner:
    \begin{eqnarray}
        F(\Lambda_{\mathfrak{s}})
        &=&
        \frac{1}{2^2}\sum_{i=1}^N\sum_{s\in\mathfrak{s}}\mathrm{Tr}\left[
         G^{-\frac{1}{2}}
         g^{(i,s)}
         G^{-\frac{1}{2}}
         g^{(i,s)}
        \right]
        \nonumber
        \\
        &=&
        F(\Lambda_{\mathfrak{s}'}),
    \end{eqnarray}
where $\Lambda_{\mathfrak{s}'}$ signifies the channel over the set of Bell outcome $\mathfrak{s}'$. This calculation mirrors the method outlined in Eq. (\ref{eq:fid_meas_ch}), with the primary distinction being the use of a unitary operator specifically designed to convert the set of Bell outcome (equivalently, the corresponding subset of Pauli operator numbers) from $\mathfrak{s}$ to $\mathfrak{s}'$. Significantly, this transformation applies solely to signal sets of identical size, owing to the requirement for one-to-one correspondence.
Therefore, we categorize PBQCT protocols according to the size of Bell outcomes, which can be $1, 2, 3,$ or $4$.  Protocols within the same category exhibit identical entanglement and teleportation fidelity. However, the distinction arises from the joint measurement and quantum operation required, which varies depending on the signal set's structure.

PBQCT protocols with a single set of signal states corresponding to a port include PBT.
Consequently, all protocols within this category can be conceptualized as simple variations of PBT.
Similarly, the protocol where the signal set is maximized is also straightforward.
In this category, the signal sum operator $ G $ effectively becomes an identity operator, making the POVM of the joint measurement identical to the signal set.
Given that the signal set can be divided by Bell states acting on different ports, the protocol effectively functions as multiple ST protocols operating in parallel. This protocol is henceforth referred to as parallel ST.
The protocols of nontrivial involve signal sets of two or three corresponding to a port, and will be referred to as PBQCT-2 and PBQCT-3, respectively.
Exhibiting a characteristic shared with PBT, these protocols demonstrate asymptotic achievement of perfect entanglement fidelity with an infinite number of ports.

In the rest of this section, we investigate PBQCT-2 and PBQCT-3, respectively.
Given that both protocols have potential applications and can be analytically expressed, we provide the analytic form of the joint POVM measurement. Furthermore, we calculate the entanglement fidelity and demonstrate its convergence towards one.

\subsection{
    \label{sec:pbqct_2}
    PBQCT-2
    }
% PBQCT-2의 정확도를 계산하기 위해, 우리는 다음과 같은 벨 넘버 아웃컴 셋을 고려한다.

We have established that PBQCT protocols with the same size of the outcome set exhibit equivalent entanglement fidelity. Building on this finding, we now focus on PBQCT-2 protocol, which utilizes a signal set defined as the set $\mathfrak{s}_2$ of Bell outcomes, given by
    \begin{eqnarray}
        \mathfrak{s}_2:=\{0,3\}.
    \end{eqnarray}
By exploiting Eq. (\ref{def:gen_srm}), we can obtain the signal sum operator $ G $ as
    \begin{eqnarray}
         G 
        &=&
        \sum^{N}_{i=1}
        \sum_{s\in\mathfrak{s}}
             g^{(i,s)}
        \nonumber
        \\
       &=&\frac{1}{2^{N}}
       \sum^{N}_{i=1}
        \Bigl[
            \left(
             \sigma^0_{A_0}\otimes \sigma^0_{A_i}
            + \sigma^3_{A_0}\otimes \sigma^3_{A_i}
            \right)
        \otimes
         I_{\bar{A}_i}
        \Bigr]
        \nonumber
        \\
        &=&\frac{1}{2^{N}}
        \Bigl[
            N
             I_{A_0\vec{A}}
            +\sum^{N}_{i=1}
             \sigma^3_{A_0}\otimes \sigma^3_{A_i}
        \otimes
         I_{\bar{A}_i}
        \Bigr].
    \end{eqnarray}
Given that the computational basis serves as its eigenbasis, we denote the eigenstates as follows:
    \begin{eqnarray}
        \ket{z_0;\vec{z}}
        :=\bigotimes_{i=0}^{N}\ket{z_i},
    \end{eqnarray}
where $\vec{z}:=\{z_1,z_2,...,z_N\}$.
We can obtain the eigenvalues of each eigenstates as follows:
    \begin{eqnarray}
         G \ket{z_0;\vec{z}}
%        &=\frac{1}{2^N}
%        \left(
%            N+(-1)^{z_0}\sum_{j=1}^{N}(-1)^{\vec{e}_j\vec{z}}
%        \right)
%        \ket{z_0;\vec{z}}
%        \nonumber\\
        &=&\frac{1}{2^N}
        \left(
            N+(-1)^{z_0}\sum_{i=1}^{N}(-1)^{z_i}
        \right)
        \ket{z_0;\vec{z}}
        \nonumber\\
        &=&\frac{1}{2^N}
        \left(
            N+(1-2z_0)\sum_{i=1}^{N}(1-2z_i)
        \right)
        \ket{z_0;\vec{z}}
        \nonumber\\
%        &=&\frac{1}{2^N}
%        \left(
%            2N(1-z_0)-2(1-2z_0)N_{\vec{z}}
%        \right)
%        \ket{z_0;\vec{z}}
        &=&\frac{c(z_0;\vec{z})^{-2}}{2^{N-1}}
        \ket{z_0;\vec{z}},
    \end{eqnarray}
where $c(z_0;\vec{z})$ is
    \begin{eqnarray}
        c(0,\vec{z})&:=&(N-N_{\vec{z}})^{-\frac{1}{2}},
    \end{eqnarray}
and
    \begin{eqnarray}
        c(1,\vec{z})&:=&N_{\vec{z}}^{-\frac{1}{2}},
    \end{eqnarray}
with
    \begin{eqnarray}
        N_{\vec{z}}:=\sum_{i=1}^Nz_i.
    \end{eqnarray}
Intuitive to the above equation, the signal sum operator $ G $ is projection operator to the null space of $\ket{0}\otimes\ket{1}^{\otimes N}$ and $\ket{1}\otimes\ket{0}^{\otimes N}$. Furthermore, the POVM elements $ \Pi^{(1,s)}$ over outcome $(1,s)$ is directly derived as follows:
    \begin{eqnarray}
         \Pi^{(1,s)}
        &=& G^{-1/2} g^{(1,s)}  G^{-1/2}+ \Delta 
        \nonumber
        \\
        &=&\sum_{\bar{z}_1}
        \left[
         G^{-1/2}
        \Bigl(\ket{\Psi^s}\otimes\ket{\bar{z}_1}\Bigr)
        \right]
        \nonumber
        \\
        & &
        \quad\;\;\;
        \cdot
        \left[
        \Bigl(\bra{\Psi^s}\otimes\bra{\bar{z}_1}\Bigr)
         G^{-1/2}
        \right]
        + \Delta 
        \nonumber
        \\
        &=&
        \sum_{\bar{z}_1}
        \ket{\Psi^s(N_{\bar{z}_1})}
        \bra{\Psi^s(N_{\bar{z}_1})}
        \otimes
        \ket{\bar{z}_1}\bra{\bar{z}_1}
        + \Delta 
        \nonumber
        \\
        &=&
        \sum_{n=0}^{N-1}
        \ket{\Psi^s(n)}
        \bra{\Psi^s(n)}
        \otimes
         I (n)
        + \Delta ,
        \label{eq:pbqct_2_srm}
    \end{eqnarray}
where $N_{\bar{z}_1}:=\sum_{i=2}^Nz_i$, and
    \begin{eqnarray}
        \ket{\Psi^{0(3)}(n)}
        &=&
        \frac{1}{\sqrt{2}}\left(\frac{1}{\sqrt{N-n}}\ket{00}\pm\frac{1}{\sqrt{1+n}}\ket{11}\right),        
    \end{eqnarray}
which describes a state that is not normalized, where $ I (n)$ represents the projector onto the subspace spanned by $N-1$ qubits, with the total number of qubits ranging from $0$ to $N-1$. It should be noted that the SRM elements for PBQCT-2 protocols, which utilize different signal sets, are derived by applying appropriate unitary transformations, as specified in Eqs. (\ref{eq:pbqct_qb_pauli_trans}) and (\ref{eq:qb_sgl_trns_last}), to the SRM elements outlined in Eq. (\ref{eq:pbqct_2_srm}).

In contrast to the PBT, which utilizes complex joint SRM elements block diagonalized based on the total spin quantum number, the implementation of measurements in PBQCT-2 is significantly more straightforward and efficient.
For PBQCT-2 protocol, SRM elements are block diagonalized solely according to the total computational number.
Furthermore, the measurement elements in PBQCT-2 can be succinctly described as Bell states acted upon by a $z$-rotation operator at the qubit $A_0$, depending on the corresponding block.
This simplicity offers a stark contrast to the intricate measurement setup required by the PBT protocol.

%%%%%%%%%%%%%%%%%%%%%%%%%%%Figure
\begin{figure*}[ht]
	\centering
	\begin{subfigure}{0.48\linewidth}
		\centering
		\includegraphics[width=\textwidth]{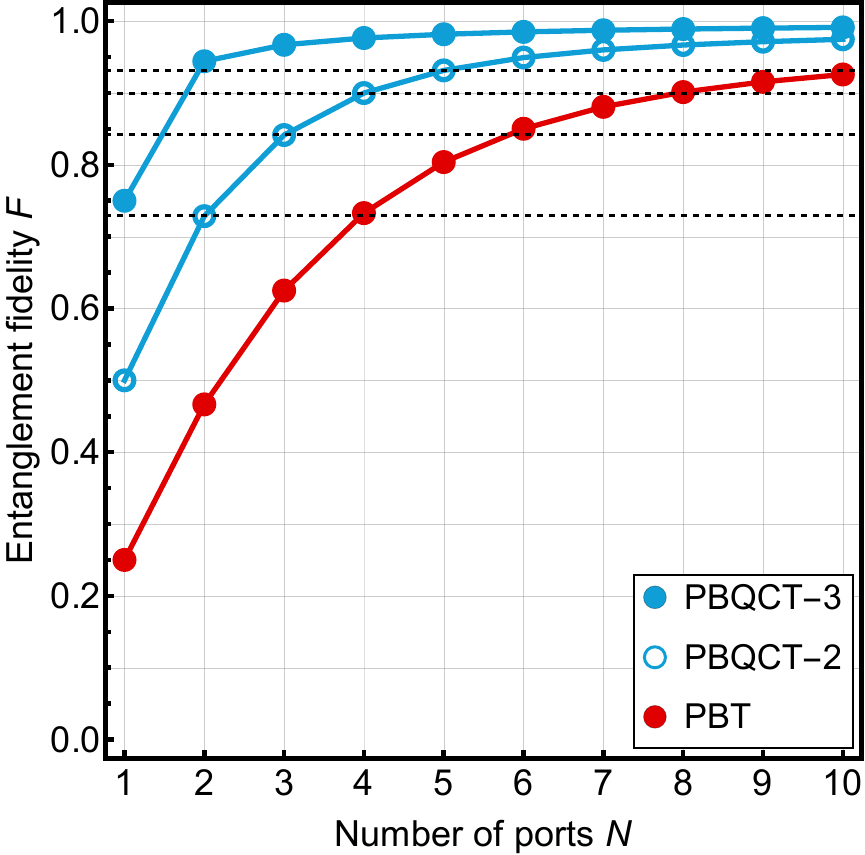} 
		\caption{}
		\label{fig:PL_entF_SMALL_2dim}
	\end{subfigure}
	\hfil
	\begin{subfigure}{0.48\linewidth}
		\centering
		\includegraphics[width=\textwidth]{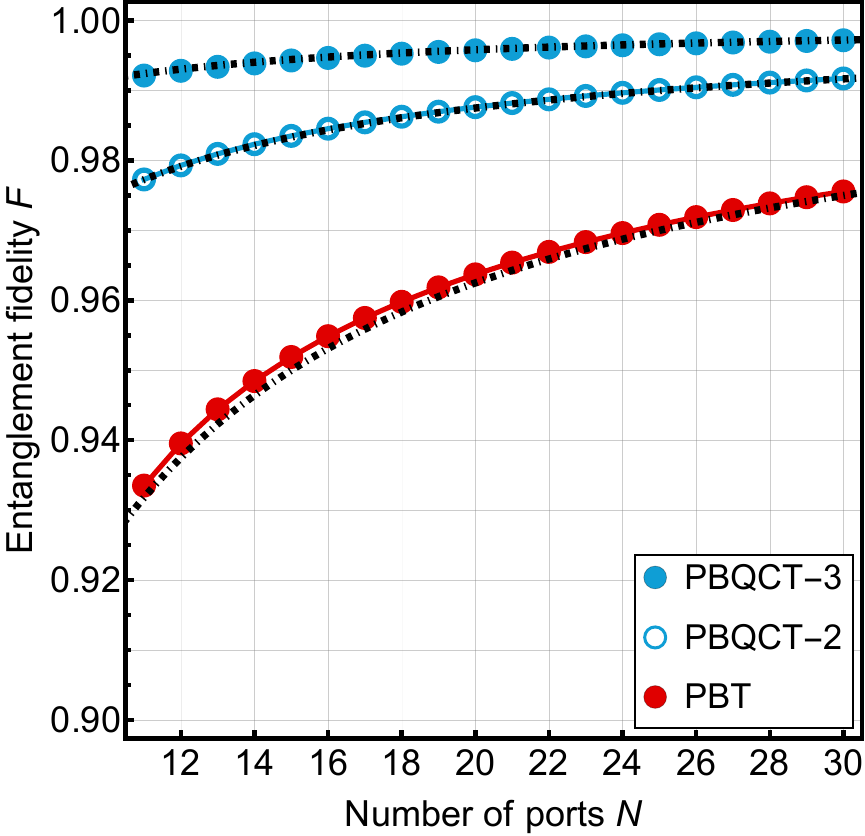}
		\caption{}
		\label{fig:PL_entF_LARGE_2dim}
	\end{subfigure}
	\caption{
        \label{figs:PL_entF_2dim}
        Entanglement fidelity $F$ for the two-dimensional PBQCT scheme, invariant under port permutations, is plotted as a function of the number of ports $N$.
        The protocols within the scheme is categorized into four group: PBT, PBQCT-2, PBQCT-3, and parallel ST, each distinguished by the size of the signal set utilized for SRM in joint measurements.
        Red closed dots represent the fidelity of the PBT protocol, blue open dots for PBQCT-2, and blue closed dots for PBQCT-3.
        The fidelity of parallel ST is not depicted as it remains constant at one regardless of $N$.
        In (a), the entanglement fidelity $F$ is displayed for a small number of ports, ranging from $N=1$ to $10$.
        The dashed lines indicate the points where the fidelity of PBQCT-2 with $N=2,3,4,5$ aligns with PBT's at half the number of ports.
        In (b), graph is extended to a larger number of ports, ranging from $N=11$ to $30$. 
        The dotted-dashed lines demonstrate the asymptotic trend of fidelity towards 1 for each protocol, conforming to the relationship $1-1/(aN)$. 
        The variable $a$ varies according to the protocol: $4/3$ for PBT, $4$ for PBQCT-2, and $12$ for PBQCT-3, illustrating each protocol's asymptotic efficiency.
        }
\end{figure*}

By exploiting Eq. (\ref{eq:def_ent_fid}), we can obtain the entanglement fidelity for the channel $\Lambda_2$ of PBQCT-2 as
    \begin{eqnarray}
        F(\Lambda_2)
        &=&\frac{N}{2}\Tr
        \left[
         \Pi^{(1,0)}
         g^{(1,0)}
        \right]
        \nonumber\\
        &=&\frac{N}{2^{N+1}}
        \sum^{N-1}_{n=0}
        |\braket{\Psi^0}{\Psi^0(n)}|^2
        \nonumber\\
        &=&\frac{N}{2^{N+2}}
        \nonumber
        \\
        & &\times
        \sum_{n=0}^{N-1}
            \left(
            \frac{1}{\sqrt{N-n}}
            +\frac{1}{\sqrt{1+n}}
            \right)^2
            {
            \mathrm{\scalebox{.7}{$N-1$}}
            \choose
            n
            }.
    \end{eqnarray}
The entanglement fidelity $F(\Lambda_2)$ is depicted using blue open dots as a function of $N$, ranging from $1$ to $10$, in Fig. \ref{fig:PL_entF_SMALL_2dim}.
Compared to the entanglement fidelity of PBT, indicated by red dots, the PBQCT-2 protocol requires half the number of ports to achieve similar fidelity.
The dashed lines in Fig. \ref{fig:PL_entF_SMALL_2dim} show that the entanglement fidelity of PBQCT-2 at $N=2,3,4,5$ approximately coincides with that of PBT at $N=4,6,8,10$, respectively.
Furthermore, for protocols with a large number of ports that bring the fidelity close to 1, the relative ratio of the number of ports required by PBQCT-2 to achieve the same fidelity as PBT decreases. 
Figure \ref{fig:PL_entF_LARGE_2dim} shows the entanglement fidelity of PBT and PBQCT-2 for port numbers from 11 to 30 with red closed and blue open dots, respectively. 
Ishizaka \textit{et al.} \cite{ishizaka2008asymptotic} demonstrated that the entanglement fidelity of PBT approaches $1-3/(4N)$ as the number of ports $N$ increases significantly. 
This is illustrated in Fig. \ref{fig:PL_entF_LARGE_2dim} as the dotted-dashed line near red dots. 
We confirmed that the fidelity function for the number of ports in PBQCT-2 has a similar tendency to that of PBT.
Compared to PBT, fidelity for PBQCT-2 approaches 1 more quickly with a smaller number of ports, as illustrated by the trend obtained through numerical fitting:
    \begin{eqnarray}
    \label{eq:pbqct_2_asym_F}
        F(\Lambda_{2})\rightarrow1-\frac{1}{4N},
    \end{eqnarray}
depicted by the dotted-dashed line near the blue open dots.
In other words, when $N$ is sufficiently large, PBQCT-2 performs equivalently to PBT with three times as many ports.

%%%%%%%%%%%%%%%%%%%%%%%%%%%%%%%%%%%%%%%%%%%%%%%%%%%%%%%%%%%%%%%%%%%%
PBQCT-2 scheme is anticipated to offer a wide range of potential applications, leveraging the SRM configuration used within the protocol and its superior performance compared to PBT.
This scheme shares many similarities with the deterministic teleportation scheme through the KLM protocol \cite{franson2002high}, which overcomes the limitations of linear optics.
Firstly, both approaches necessitate sequential classical and quantum corrections following measurement, albeit in different sequences.
Secondly, they share same null space, impacting entanglement fidelity.
Third, without optimization of resource states, fidelity diminishes inversely proportional to the number of ports.
Thus, these similarities between the two schemes provide a novel angle to reinterpret the KLM protocol from a fresh perspective and, conversely, open up possibilities for implementing the PBQCT-2 scheme using linear optics.

For other application, PBQCT-2 provides the prospect of enhanced performance or error-protected teleportation protocols that do not necessitate quantum correction by the receiver, similar to PBT. 
In the PBQCT-2 scenario, Bob can accurately receive the unknown state by either applying the Pauli $z$ operator to a specified port or leaving it unchanged, based on the outcome communicated by Alice.
Given that the receiver corrections commute with each other, the process can be described in the stabilizer formalism.
Accordingly, if the state information is encoded to be resilient against alterations following measurements, it enables the transmission of states without necessitating any quantum corrections during the process.
Building upon PBQCT-2 and the stabilizer formalism, we are presently exploring alternative PBT protocols.
    
\subsection{
    PBQCT-3
    }

In a manner analogous to our examination of PBQCT-2, we now explore PBQCT-3, characterized by a signal set defined as set $\mathfrak{s}_3$ of Bell outcome, given by
    \begin{eqnarray}
        \mathfrak{s}_3:=\{0,1,3\}.
    \end{eqnarray}
Subsequently, we derive the signal sum operator $ G $ as follows:
    \begin{eqnarray}
         G 
        &=&
        \sum^{N}_{j=1}
        \sum_{s\in\mathfrak{s}}
             g^{(j,s)}
        \nonumber
        \\
        &=&
        \sum^{N}_{j=1}
        \left(
            \frac{1}{2^{N-1}} I - g^{(j,2)}
        \right)
        \nonumber
        \\
        &=&
        \frac{N}{2^{N-1}} I - G_1,
    \end{eqnarray}
where $ G_1$ is the signal sum operator of PBT, considering maximally entangled states as spin singlets.
Since $ G $ can be expressed as the summation of the identity operator and $ G_1$, it shares the same eigenstates as $ G_1$.
Thus, the eigenvalue equation of $ G $ is given as:
    \begin{eqnarray}
         G_{\mp}\ket{\Psi^{[N]}_\mp
        (s,s_z;\alpha)}=\lambda^\mp_{s}\ket{\Psi^{[N]}_\mp(s,s_z;\alpha)},
    \end{eqnarray}
where $ G := G_{-}\oplus G_{+}$, $\lambda^{-}_{s}:=(3N/2+s)/2^N$, and $\lambda^{+}_{s}:=(3N/2-s+1)/2^N$.
Additionally, $s$ represents the spin angular momentum, ranging from $0(1/2)$ to $N/2$ for even(odd) $N$, $s_z$ runs from $-s$ to $s$, and $\alpha$ specifies the additional degree of freedom.
The eigenstates are represented using the orthogonal basis ${\ket{s,s_z;\alpha}}$ of the $N$-spin system, i.e., the basis of the irreducible representation of $SU(2)^{\otimes N}$. 
The computational basis expression can also be obtained using the recursive relation within the basis. For further details, refer to the appendix in the preprint version of \cite{ishizaka2008asymptotic}.
By representing identity with the basis ${\ket{s,s_z;\alpha}}$, signal states can be denoted as:
    \begin{eqnarray}
         g^{(i,r)}_{A_0\vec{A}}
        &=&
        \frac{1}{2^{N-1}}
        \sum_{s_\mathrm{min}}^{(N-1)/2}
        \sum_{s_z,\alpha}\ket{\Psi^r}\bra{\Psi^r}_{A_0A_i}
        \nonumber
        \\
        & &\qquad\qquad\qquad\qquad\otimes
        \ket{s,s_z;\alpha}\bra{s,s_z;\alpha}_{\bar{A}_i}.
    \end{eqnarray}
We can obtain the entanglement fidelity for the channel $\Lambda_3$ of PBQCT-3 as
    \begin{eqnarray}
        F(\Lambda_3)
        &=&\frac{3N}{4}\Tr
        \left[
         \Pi^{(1,0)}
         g^{(1,0)}
        \right]
        \nonumber\\
        &=&\frac{3N}{4}\Tr
        \left[
         G^{-\frac{1}{2}}
         g^{(1,0)}
         G^{-\frac{1}{2}}
         g^{(1,0)}
        \right]
        \nonumber\\
        &=&\frac{3N}{4}
        \sum_{s=s_\mathrm{min}}^{(N-1)/2}
        \sum_{s_z,\alpha}
        \left[c_+(s,s_z)+c_-(s,s_z)\right]^2
        \nonumber\\
        &=&\frac{3N}{4}
        \sum_{s,s_z}
        \frac{(2s+1)(n-1)!}{((n-1)/2-s)!((n+1)/2+s)!}
        \nonumber
        \\
        & &\qquad\quad\;\;\times
        \left[c_+(s,s_z)+c_-(s,s_z)\right]^2
        ,
    \end{eqnarray}
where $c_+(s,s_z)$ and $c_-(s,s_z)$ are as follows:
    \begin{widetext}
    \begin{eqnarray}
        c_-(s,s_z)
        &=&\left(
        \bra{\Psi^1}\otimes\bra{s,s_z;\alpha}
        \right)
         G^{-\frac{1}{2}}_{-}
        \left(
        \ket{\Psi^1}\otimes\ket{s,s_z;\alpha}
        \right)
        \nonumber\\
        &=&\frac{1}{2}
        \left(\lambda^-_{s_+}\right)^{-\frac{1}{2}}
        |
        \cg{s_+}{s_{z+}}{s_{++}}_-
        \cg{s}{s_{z}}{s_+}_+
        +
        \cg{s_+}{s_{z-}}{s_{++}}_+
        \cg{s}{s_{z}}{s_+}_-
        |^2
        \nonumber\\
        & &+\frac{1}{2}
        \left(\lambda^-_{s_-}\right)^{-\frac{1}{2}}
        |
        \cg{s_-}{s_{z+}}{s}_-
        \cg{s}{s_{z}}{s_-}_+
        +
        \cg{s_-}{s_{z-}}{s}_+
        \cg{s}{s_{z}}{s_-}_-
        |^2
        \nonumber\\
        &=&
        \left(\lambda^-_{s_-}\right)^{-\frac{1}{2}}
        \frac{s_z^2}{s(1+2s)}
        +\left(\lambda^-_{s_+}\right)^{-\frac{1}{2}}
        \frac{(1+s)^2-s_z^2}{(1+s)(1+2s)},
    \end{eqnarray}
and
    \begin{eqnarray}
        c_+(s,s_z)
        &=&\left(
        \bra{\Psi^1}\otimes\bra{s,s_z;\alpha}
        \right)
         G^{-\frac{1}{2}}_{+}
        \left(
        \ket{\Psi^1}\otimes\ket{s,s_z;\alpha}
        \right)
        \nonumber\\
        &=&\frac{1}{2}
        \left(\lambda^+_{s_-}\right)^{-\frac{1}{2}}
        |
        \cg{s_-}{s_{z+}}{s_{--}}_-
        \cg{s}{s_{z}}{s_-}_+
        +
        \cg{s_-}{s_{z-}}{s_{--}}_+
        \cg{s}{s_{z}}{s_-}_-
        |^2
        \nonumber\\
        & &+\frac{1}{2}
        \left(\lambda^+_{s_+}\right)^{-\frac{1}{2}}
        |
        \cg{s_+}{s_{z+}}{s}_-
        \cg{s}{s_{z}}{s_+}_+
        +
        \cg{s_+}{s_{z-}}{s}_+
        \cg{s}{s_{z}}{s_+}_-
        |^2
        \nonumber\\
        &=&
        \left(\lambda^+_{s_-}\right)^{-\frac{1}{2}}
        \frac{s^2-s_z^2}{s(1+2s)}
        +\left(\lambda^+_{s_+}\right)^{-\frac{1}{2}}
        \frac{s_z^2}{(1+s)(1+2s)}.
    \end{eqnarray}
    \end{widetext}
Here, we introduced a shorthand notation for (nonvanishing) Clebsch-Gordan coefficients,
    \begin{eqnarray}
        \cg{j_1}{m_1}{j}
        :=
        \left\langle j_1,m_1,\frac{1}{2}\pm\frac{1}{2}
        \;\Biggl|\;
        j,m_1\pm\frac{1}{2}
        \right\rangle
    \end{eqnarray}
and write $s_\pm=s\pm1/2$, $s_{\pm\pm}=s\pm1$, and $s_{z\pm}=s_z\pm1/2$.
The corresponding entanglement fidelity $F(\Lambda_3)$, depicted by blue closed dots, varies as a function of $N$ from $1$ to $10$ as shown in Fig. \ref{fig:PL_entF_SMALL_2dim}, and from $11$ to $30$ in Fig. \ref{fig:PL_entF_LARGE_2dim}. 
Figure \ref{fig:PL_entF_SMALL_2dim} reveals that the entanglement fidelity surpasses $0.9$ even with a two-port protocol.
Even with a small number of ports, the entanglement fidelity approximates to     \begin{eqnarray}
    \label{eq:pbqct_3_asym_F}
        F(\Lambda_{3})\rightarrow1-1/(12N),
    \end{eqnarray}
as depicted by the dotted-dashed line adjacent to the blue closed dots.
When compared to PBT, PBQCT-3 performs equivalently to PBT protocols that utilize nine times as many ports.

\section{Qudit systems}
\label{sec:high}

In a similar manner to qubit systems, we restrict PBQCT in qudit systems ($D>2$) to have signal states in the form of:
   \begin{eqnarray}
   \label{def:qd_sgnsts}
         g^{(i;x,y)}
        :=\frac{1}{D^{N-1}} \Psi^{(x,y)}_{A_0A_i}
        \otimes
         I_{\bar{A}_i},
    \end{eqnarray}
where $ \Psi^{(x,y)}$ is defined as a generalized Bell state in Eq. (\ref{eq:ST_POVM_sts}).
Using the same assumption as in qubit systems, we consider PBQCTs with outcome sets given by
    \begin{eqnarray}
    \label{def:qd_outcome_set}
        \mathfrak{m}:=
        \{
            (i;x,y)\;:\;
            i\in\{1,...,N\},\;(x,y)\in\mathfrak{p}
        \},
    \end{eqnarray}
where $\mathfrak{p}$ is a nonempty subset of generalized Bell states' numbers $\{(x,y):x,y\in\{0,1,...,D-1\}\}$.
%%%%%%%%%%%%%%%%%%%%%%%%%%%Figure
    \begin{figure}[h]
	\centering
        \includegraphics[width=\linewidth]{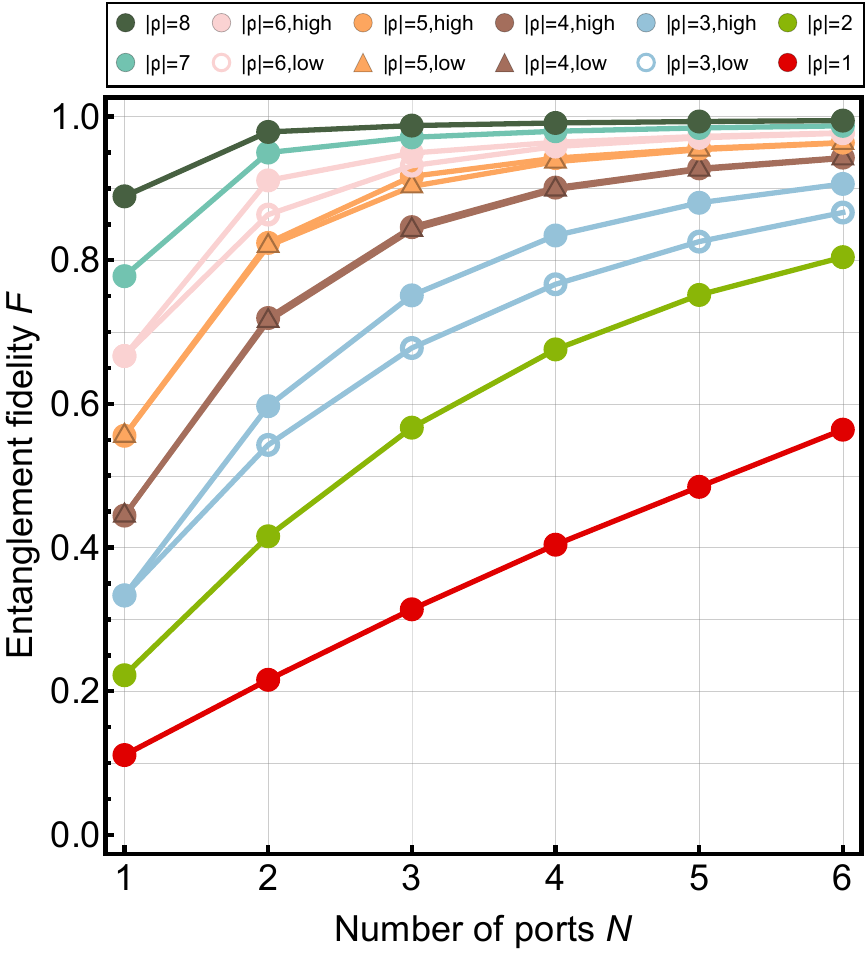}
        \caption{\label{figs:PL_entF_Ddim}
        Entanglement fidelity $F$ for the three-dimensional PBQCT scheme, which remains invariant under port permutations, is depicted as a function of the number of ports $N$.
        Each marker represents protocols with different sets of generalized Bell outcomes $\mathfrak{p}$.
        Unlike qubit systems, varying fidelities are observed even within protocols having the same size of generalized Bell outcome sets.
        The baby blue, brown, orange, and light pink lines correspond to fidelity of protocols with sizes of $\mathfrak{p}$ equal to $3$, $4$, $5$, and $6$, respectively.
        Given that the number of ports $N$ increases, the protocols divide into two categories: high fidelity and low fidelity, each denoted by markers of distinct shapes.}
    \end{figure}

The identical entanglement fidelity of measurement and teleportation channels in qubit systems stems from the presence of permutation-invariant unitary operators. These operators facilitate the transformation described in Eq. (\ref{eq:pbqct_qb_pauli_trans}).
This fundamental principle ensures that, despite differences in outcome sets, all protocols within the same size of set exhibit the same fidelity.
Unlike in qubit systems, finding a transformation between signal sets of the same size in qudit systems is not straightforward and impossible in most cases.
In Fig. \ref{figs:PL_entF_Ddim}, we present the entanglement fidelity for PBQCT in three-dimensional systems, tracking its changes from $N=1$ to $N=6$.
Each line with markers, distinguished by its own color, represents the entanglement fidelity associated with a specific size of the signal set $\mathfrak{p}$.
Notably, the lines in baby blue, brown, orange, and light pink correspond to PBQCT protocols with $\mathfrak{p}$ sizes of 3, 4, 5, and 6, respectively.
Here, we observe a divergence in entanglement fidelity, indicating two distinct types of behavior.
This distinction highlights that the transformation detailed in Eq. (\ref{eq:pbqct_qb_pauli_trans}), when applied to subsets of Weyl-Heisenberg operators of the same size, does not apply universally across systems in higher dimensions.
Additionally, the lines in Fig. \ref{figs:PL_entF_Ddim} representing the entanglement fidelity for different sizes or types of the signal set $\mathfrak{p}$ do not intersect.
This consistent ordering demonstrates that protocols with smaller sizes of $\mathfrak{p}$ consistently exhibit lower fidelity across the range of port numbers, and this hierarchy remains unchanged as the number of ports increases.

In the next subsection, we delve into two different region of PBQCT scheme.
Initially, we show that every protocol under the PBQCT scheme guarantees perfect teleportation with an infinite number of ports.
Following that, we generalize the PBQCT-2 scheme from qubit systems to higher-dimensional.
Here, we derive the analytical expression of joint POVM elements and evaluate entanglement fidelity, showcasing PBQCT's adaptability and efficiency across various dimensional settings.

\subsection{Asymptotically perfect fidelity}
\label{subsec:asymptotic}

We demonstrate that within the PBQCT scheme, perfect teleportation fidelity is attainable in the asymptotic limit across all protocols.
Consider $\mathfrak{m}$ as the total outcome set and $\mathfrak{p}$ as the individual port outcome set for a PBQCT encompassing $N$ ports, where the relationship $|\mathfrak{m}|=N|\mathfrak{p}|$ holds as per definition (\ref{def:qd_outcome_set}).
The signal states are defined in Eq. (\ref{def:qd_sgnsts}).
When the joint measurement projects the prepared state to one of the signal states, Bob can perfectly receive the unknown state by selecting the appropriate port and applying the corresponding Weyl-Heisenberg operator.
Unfortunately, the signal set alone does not constitute a complete POVM, requiring additional considerations.
To address this, deterministic PBQCT protocols employ the signal sum operator to establish a joint POVM.
The definition of SRM, as denoted in Eq. (\ref{def:gen_srm}), guarantees that the closer the signal sum operator $ G $ is to the identity operator, the higher the fidelity of transmission.
For this reason, we begin our discussion by examining the distance between the signal sum operator $ G $ and the identity operator given by
    \begin{eqnarray}
        \left\| G -\alpha  I 
        \right\|_2
        &=&
        \mathrm{Tr}
        \left[
        \left( G -\alpha I \right)
        \left( G^\dagger-\alpha I \right)
        \right]
        \nonumber
        \\
        &=&\mathrm{Tr}
        \left[
         G^2
        \right]
        -2\alpha\mathrm{Tr}
        \left[
         G 
        \right]
        +\alpha^2\mathrm{Tr}
        \left[ I 
        \right]
        \nonumber
        \\
        &=&\frac{|\mathfrak{m}|}{D^{N+1}}
        \left(
        D^2+|\mathfrak{m}|-\frac{|\mathfrak{m}|}{N}
        \right)
        \nonumber\\
        & &
        -2\alpha
        |\mathfrak{m}|
        +\alpha^2D^{N+1},
        \label{eq:qd_dist_3}
    \end{eqnarray}
where $\|\cdot\|_2$ denotes the Hilbert–Schmidt norm, and $\alpha$ a is proportional constant associated with the measurement probability.
In Eq. (\ref{eq:qd_dist_3}), the trace of operators is calculated using Eqs. (\ref{eq:ST_POVM_sts}) and (\ref{def:qd_sgnsts}).
By differentiating Eq. (\ref{eq:qd_dist_3}) with respect to $\alpha$, 
we obtain $\alpha^*$ as
    \begin{eqnarray}
        \alpha^*
        =\frac{|\mathfrak{m}|}{D^{N+1}}
    \end{eqnarray}
that minimizes the distance given by
    \begin{eqnarray}
        \left\| G -\alpha^*  I 
        \right\|_2
        &=&\frac{N}
        {D^{N+1}}
        \left(D^2-|\mathfrak{p}|\right)
        |\mathfrak{p}|.
        \label{eq:qd_dist_min}
    \end{eqnarray}
The minimal distance shows that $ G $ exponentially approaches $\alpha^* I $ as the number of ports increases. Additionally, the distances with individual port result set sizes as $|\mathfrak{p}|$ and $D^2-|\mathfrak{p}|$ are equivalent.

We define the operator $ O $ as
    \begin{eqnarray}
         O := G -\alpha^* I ,
    \nonumber
    \end{eqnarray}
where it becomes the zero operator in the asymptotic limit $N\rightarrow\infty$.
By exploiting (\ref{eq:def_ent_fid}), we derive the entanglement fidelity for the PBQCT channel $\Lambda_\mathfrak{m}$ as
    \begin{eqnarray}
        F(\Lambda_\mathfrak{m})
        &=&
        \frac{1}{D^2}\sum_{m\in\mathfrak{m}}\mathrm{Tr}
        \Bigl[
        ( O +\alpha^* I )^{-\frac{1}{2}}
         g^m
        \nonumber\\
        & &\qquad\qquad\quad\;\cdot
        ( O +\alpha^* I )^{-\frac{1}{2}}
         g^m
        \Bigr]
%        \\
%        &\rightarrow
%       \frac{|\mathfrak{m}|}{D^2\alpha^*}
%        \mathrm{Tr}
%        \left[
%             G^2
%        \right]
%        -\frac{1}{D^2}
%       \sum_{m\in\mathfrak{m}}
%        \mathrm{Tr}\left[ g^m O  g^m\right]
        \nonumber\\
        &\rightarrow&
        1
        -\frac{1}{D^2}
        \sum_{m\in\mathfrak{m}}
        \mathrm{Tr}\left[ g^m O  g^m\right]
        ,
        \label{eq:qd_asym_ent_fid}
    \end{eqnarray}
where we disregard traces of operators containing two or more instances of $ O $, given that $ O $ approaches the zero operator in the asymptotic limit.
Given that the last term of Eq. (\ref{eq:qd_asym_ent_fid}) vanishes in $N\rightarrow\infty$, we deduce that all PBQCTs achieve asymptotically perfect entanglement fidelity, meaning PBQCT protocols perfectly transmit the unknown state in infinite number of ports.
Remarkably, this conclusion holds true even for PBQCT protocols with broken permutation symmetry in the signal set, as the only variance lies in the value of $\mathrm{Tr}[ G^2]$, with all other aspects remaining unchanged.

\subsection{Generalized PBQCT-2}
\label{subsec:gen_pbqct_2}

    \begin{figure}[h]
	\centering
        \includegraphics[width=\linewidth]{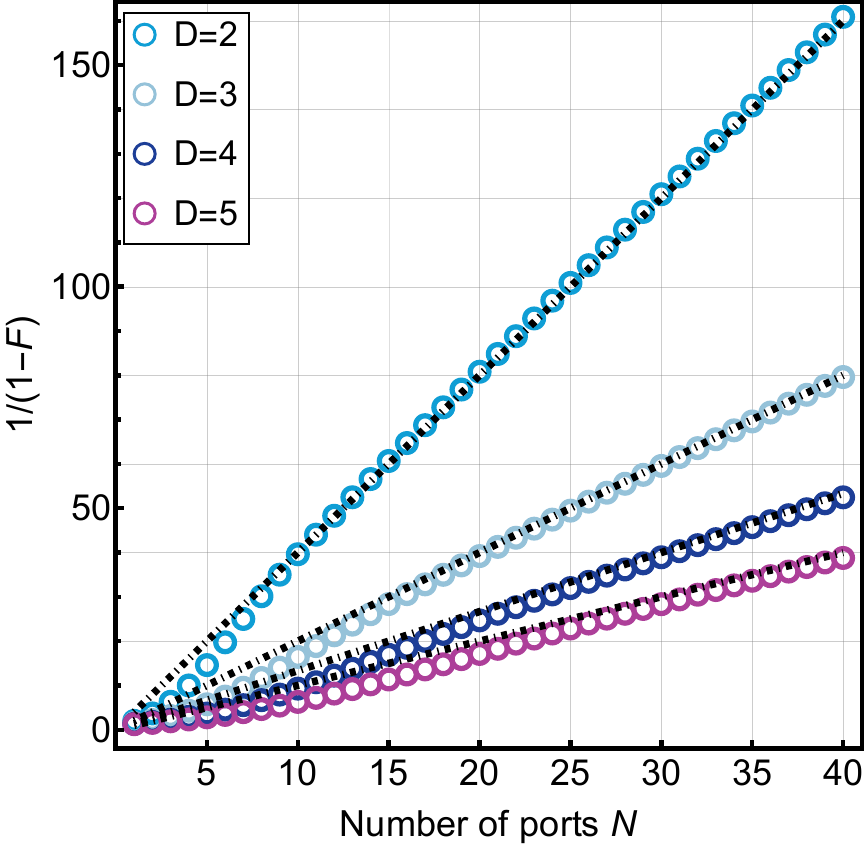}
        \caption{\label{figs:PL_entF_LARGE_Ddim}
        The reformulated entanglement fidelity for $D$-dimensional generalized PBQCT-2, with the size of generalized Bell outcomes set equals to $D$, is shown as $1/(1-F)$ and plotted against the number of ports $N$.
        The dotted-dashed lines indicate the asymptotic trend of fidelity approaching $1$ for each dimension $D=2,3,4,5$, adhering to the relationship $F\rightarrow1-(D-1)/(4N)$.}
    \end{figure}

PBQCT protocols exhibiting properties akin to those observed in the PBQCT-2, as discussed in section \ref{sec:pbqct_2}, are also present in higher-dimensional systems.
A defining characteristic of the PBQCT-2 protocol comes from its signal sum operator, which has eigenbasis as the computational basis.
This characteristic persists in protocols where each port encompasses either all or none of the outcomes linked to the exponent of the shift operator, as indicated in Eq. (\ref{def:shft_mat}).
Considering a signal set defined in Eq. (\ref{def:qd_sgnsts}), we introduce an individual port outcome set $\mathfrak{p}_{\mathfrak{q}}$, as specified by
    \begin{eqnarray}
        \mathfrak{p}_{\mathfrak{q}}=
            \left\{
            (x,y)
            \;:\;
            x\in \{0,1,...,D-1\}
            ,\;
            y\in \mathfrak{q}
            \right\},
    \end{eqnarray}
where $\mathfrak{q}$ is a non empty subset of $\{0,1,...,D-1\}$.
The signal sum operator $ G $ is obtained as follows:
    \begin{eqnarray}
         G 
        &=&
        \sum^{N}_{i=1}
        \sum_{x=0}^{D-1}
        \sum_{y\in\mathfrak{q}}
             g^{(i;x,y)}
        \\
       &=&\frac{1}{D^{N-1}}
        \sum^{N}_{i=1}
        \sum_{y\in\mathfrak{q}}
        \sum^{D-1}_{j=0}
            \ket{j}\bra{j}_{A_0}
       \nonumber\\
       & &\qquad\qquad\qquad\qquad\;    
        \otimes\ket{j\oplus y}\bra{j\oplus y}_{A_i}
        \otimes
         I_{\bar{A}_i}.
    \end{eqnarray}
We can obtain the eigenvalues of each eigenstates as follows:
    \begin{eqnarray}
         G \ket{z_0;\vec{z}}
%        &=&\frac{1}{D^{N-1}}
%       \sum^{N}_{j=1}
%       \sum_{p}^{k}
%       \sum^{D}_{i=1}
%            \braket{i}{z_0}
%            \braket{i+r}{z_j}
%        \ket{i}_{A_0}
%        \otimes    
%        \ket{i+r}_{A_j}
%        \otimes
%        \ket{\bar{z}_j}
%        \nonumber\\
        &=&\frac{1}{D^{N-1}}
       \sum^{N}_{i=1}
       \sum_{y\in\mathfrak{q}}
       \sum^{D-1}_{j=0}
            \delta_{j,z_0}
            \delta_{j\oplus y,z_i}
        \ket{j}_{A_0}
        \nonumber\\
        & &\qquad\qquad\qquad\qquad\;
        \otimes    
        \ket{j\oplus y}_{A_j}
        \otimes
        \ket{\bar{z}_i}
        \nonumber\\
%        &=&\frac{1}{D^{N-1}}
%       \sum^{N}_{j=1}
%       \sum_{p=1}^k
%            \delta_{z_0+r,z_j}
%            \ket{z_0}_{A_0}
%            \otimes
%            \ket{z_j}_{A_j}
%        \otimes
%        \ket{\bar{z}_j}
%        \nonumber\\
        &=&\frac{c(z_0;\vec{z})^{-2}}{D^{N-1}}
        \ket{z_0;\vec{z}},
    \end{eqnarray}
where $c(z_0;\vec{z})$ is
    \begin{eqnarray}
        c(z_0,\vec{z})&=&
        \left(
        \sum_{i=1}^{N}
        \sum_{y\in\mathfrak{q}}
        \delta_{z_0\oplus y,z_i}
        \right)
        ^{-\frac{1}{2}}
        \nonumber
        \\
        &=&\left(N^{z_0}_{z_1}+N^{z_0}_{\bar{z}_1}\right)^{-\frac{1}{2}}
    \end{eqnarray}
with
    \begin{eqnarray}
        N_{z_1}^{z_0}
        :=\sum_{y\in\mathfrak{q}}
        \delta_{z_0\oplus y,z_1},
    \end{eqnarray}
and
    \begin{eqnarray}
    \label{eq:num_barZ1}
        N_{\bar{z}_1}^{z_0}
        :=
        \sum_{j=2}^N
        \sum_{y\in\mathfrak{q}}
        \delta_{z_0\oplus y,z_j}
        .
    \end{eqnarray}
Whenever all elements of $\vec{z}$ are not a part of $z_0\oplus y$ for $y\in\mathfrak{q}$, the eigenstates $\ket{z_0;\vec{z}}$ has an eigenvalue of zero.
Consequently, we find that $ \Delta $ is the projection operator into the space of these eigenstates.
By exploiting Eq. (\ref{def:gen_POVM}), we obtain the POVM elements $ \Pi^{(1;p,q)}$ over first port as 
    \begin{eqnarray}
         \Pi^{(1;p,q)}
        &=& G^{-1/2} g^{(1;p,q)} G^{-1/2}+ \Delta 
        \nonumber
        \\
        &=&
        \sum_{\vec{n}\in \mathfrak{N}}
        \ket{\Psi^{(p,q)}(\vec{n})}
        \bra{\Psi^{(p,q)}(\vec{n})}
        \otimes
         I (\vec{n})
        + \Delta ,
    \label{eq:pbqct_dk_srm}
    \end{eqnarray}
where the equation represents the generalized form of Eq. (\ref{eq:pbqct_2_srm}) for a qudit systems, extending the results obtained from the qubit systems.
%Integer partitions
In Eq. (\ref{eq:pbqct_dk_srm}), $\mathfrak{N}$ denotes the set of all possible sequences of integer partitions, where each sequence has a length $D$ and contains integers ranging from $0$ to $N-1$. Formally, $\mathfrak{N}$ is defined as
    \begin{eqnarray}
        \mathfrak{N}&:=&
        \Bigl\{
        \{n_0,n_1,...,n_{D-1}\}
        \;:\;
        \sum_{i=0}^{D-1} n_i= |\mathfrak{q}|(N-1)
        \nonumber\\
        & &
        \qquad\qquad\qquad\qquad\qquad
        ,\;
        \forall n_i\in[0,N-1]
        \Bigr\},
    \end{eqnarray}
Additionally, $\ket{\Psi^{(p,q)}(\vec{n})}$ is given by
    \begin{eqnarray}
        \ket{\Psi^{(p,q)}(\vec{n})}
        &=&
        \frac{1}{\sqrt{D}}\sum_{i=0}^{D-1}\frac{e^{\frac{2\pi i}{N}jp}}{\sqrt{1+n_i}}\ket{i}\otimes{\ket{i+q}}
        \nonumber\\      
        &=&
        \bigl( W^{(p,q)}\otimes I \bigr)\Ket{\Psi^{(0,0)}(\vec{n})}.
    \end{eqnarray}
and the projector $ I (\vec{n})$ is defined over a subspace that encompasses $(N-1)$-qubit states $\ket{z_2,z_3,...,z_{N}}$, which represent specific combinations of states excluding qubit $A_0$ and $A_1$. For each index $i$ within the set $\{0,1,...,D-1\}$, $n_i$ matches $N_{\bar{z}_1}^i$, as defined in Eq. (\ref{eq:num_barZ1}).
For instance, consider the circumstance where $\vec{n} = {4,1,0}$ in a system with $D=3$ and $N=6$, and where $\mathfrak{q}={0}$.
In this case, $ I (\vec{n})$ projects onto a subspace formed by the states $\ket{10000}$, $\ket{01000}$, $\ket{00100}$, $\ket{00010}$, and $\ket{00001}$.
Similarly, for $D=2$, the notation $ I (n)$ used in Eq. (\ref{eq:pbqct_2_srm}) is equivalent to $ I ({N-1-n,n})$ in our current context.    

In particular, for the protocol utilizing $\mathfrak{q}=\{0\}$, commutative nature of receiver corrections aligns with the properties seen in PBQCT-2.
By exploiting Eq. (\ref{eq:def_ent_fid}), we can obtain the entanglement fidelity for the channel $\Lambda_{\{0\}}$ of generalized PBQCT-2 with port outcome set $\mathfrak{p}_{\{0\}}$ as
    \begin{eqnarray}
        F\left(\Lambda_{\{0\}}\right)
        &=&
        \mathrm{Tr}
        \left[
             \Psi^{(0,0)}
                \left(\Lambda_{\{0\}}
                \otimes \mathbf{1}\right)
             \Psi^{(0,0)}
        \right]
        \nonumber\\
        &=&\frac{N}{D}
        \sum_{x=0}^{D-1}
        \Tr
        \left[
         \Pi^{(1;x,0)}
         g^{(1;x,0)}
        \right]
        \nonumber\\
        &=&\frac{N}{D^{N+1}}
        \sum_{\vec{n}\in\mathfrak{N}}
        |\braket{\Psi^0}{\Psi^0(\vec{n})}|^2
        \nonumber\\
        &=&\frac{N}{D^{N+2}}
        \nonumber\\
        & &\times
        \sum_{\vec{n}\in\mathfrak{N}}
            \left(
            \sum_{i=0}^{D-1}
            \frac{1}{\sqrt{1+n_i}}
            \right)^2
            {N-1 
            \choose
            n_0
            \mathrm{\scalebox{.5}{$\cdots$}}
            n_{\mathrm{\scalebox{.6}{$D-1$}}}
            },
        \label{eq:fid_gen_pbqct_1}
    \end{eqnarray}
where the last term of Eq. (\ref{eq:fid_gen_pbqct_1}) denotes multinomial coefficient.
Figure \ref{figs:PL_entF_LARGE_Ddim} illustrates the modified parameter based on the entanglement fidelity of the generalized PBQCT-2 protocol, covering port numbers from 1 to 40 across various dimensions.
This parameter adjustment is based on the observation that fidelity nears 1 inversely with the increase in the number of ports, particularly when the number of ports is large.
We observe that the entanglement fidelity nears 1 according to the formula:
    \begin{eqnarray}
        F\left(\Lambda_{\{0\}}\right)
        \rightarrow
        1-\frac{D-1}{4N},
    \end{eqnarray}
as illustrated by the dashed lines for each dimension in Fig. \ref{figs:PL_entF_Ddim}.
On the other hands, Christandl \textit{et al.} \cite{christandl2021asymtotic} demonstrate that the entanglement fidelity of deterministic PBT approaches $F\left(\Lambda_{\mathrm{PBT}}\right)
        \rightarrow
        1-(D^2-1)/(4N)$.
It indicates that the fidelity of the generalized PBQCT-2 converges to one more swiftly than that of PBT, proportional to the system's dimensionality.
This accelerated convergence in generalized PBQCT-2 is attributed to the utilization of an additional $D-1$ measurement elements for each port.

\section{Conclusions and Discustions}

In this paper, we considered a scheme of asymptotic quantum teleportation where $N$ EPR pairs was shared, and Alice utilized SRM measurement elements induced by signal sets composed with maximally entangled states.
At the end of the process, Bob obtains the teleported state by selecting one of the $N$ ports and correcting it by unitary operation.
We referred to this scheme as PBQCT.
The features and performance of PBQCT are influenced and depend upon the composition and size of the signal set.
Under minimal size of signal set, PBQCT contains the PBT protocol, while increasing the size of signal set to the maximum converts it to the parallel ST protocol.
Our analysis through PBQCTs connects the relation between parallel ST and PBT as a function of the size of signal set.

We showed that every PBQCT protocols provides perfect entanglement fidelity in the asymptotic limit for any types and dimensions.
Specially for qubit systems, we catagorized PBQCT into four groups according to entanglement fidelity.
Importantly, protocols within the same group can be transformed with local unitary transformation at POVM and quantum correction operators.
We analytically evaluated the POVM elements and entanglement fidelity of every PBQCT protocols.
Furthermore, we numerically investigated PBQCT for qudit systems and found that the fidelities are enhanced as the size of the signal set increases. 

The PBQCT protocols with SRM induced by two signal states for each port, denoted as PBQCT-2, shows potential applicability for finding experimentally feasible teleportation schemes such that quantum correction is unnecessary.
The first proposed PBT protocol has many difficulties in implementing it because the SRM elements are diagonalized to the Shur basis.
On the other hand, PBQCT-2 appears to have greater implementation potential in that it is block diagonalized for computational total number and has a simple block form.
Additionally, given that the null space where teleportation fails is the same as the KLM protocol, it is expected that it will be possible to express it with linear optics.
Due to the necessity of Pauli correction in PBQCT-2, it cannot be classified as a protocol possessing identical functionality to PBT.
The required quantum correction can be overcome by applying a simple two-bit concatenated code.
If we apply a more generalized error code to PBQCT, we expect to be able to find protocols that can be utilized as a universal programmable processor which is also protected from various errors.

For high-dimensional applications, the generalized PBQCT-2 protocol retains all characteristics of PBQCT across higher dimensions.
This opens new avenues for creating universal programmable processors utilizing qudit systems, offering significant advancements in quantum computing.
Besides to being a valuable and effective guide in finding nonnecessary quantum correction teleportation, PBQCT promises to be a good map for classifying existing asymptotic quantum teleportation.
Just as ST and PBT have been linked through modification of POVM measurements, we expect that other teleportation protocols, like KLM protocol and catalytic teleportation, may also be classified as one of the PBQCT protocols if additional variations such as different form of resource state, LOCC, and noise is allowed.

\section*{Acknowledgements}

This work was supported by the National Research Foundation of Korea through a grant funded by the Ministry of Science and ICT (NRF-2022M3H3A1098237), the Ministry of Education (NRF-2021R1I1A1A01042199), and Korea Institute of Science and Technology Information. H.E.K. acknowledges support by Creation of the Quantum Information Science R\&D Ecosystem through the National Research Foundation of Korea funded by Ministry of Science (Grant Nos. NRF-2023R1A2C1005588 and NRF-2021M3H3A1036573).
All numerical calculations and figures were performed using Wolfram Research, Inc., Mathematica, Version 13.3, Champaign, IL (2023).

%\section*{References}

%


\begin{thebibliography}{0}%
\makeatletter
\providecommand \@ifxundefined [1]{%
 \@ifx{#1\undefined}
}%
\providecommand \@ifnum [1]{%
 \ifnum #1\expandafter \@firstoftwo
 \else \expandafter \@secondoftwo
 \fi
}%
\providecommand \@ifx [1]{%
 \ifx #1\expandafter \@firstoftwo
 \else \expandafter \@secondoftwo
 \fi
}%
\providecommand \natexlab [1]{#1}%
\providecommand \enquote  [1]{``#1''}%
\providecommand \bibnamefont  [1]{#1}%
\providecommand \bibfnamefont [1]{#1}%
\providecommand \citenamefont [1]{#1}%
\providecommand \href@noop [0]{\@secondoftwo}%
\providecommand \href [0]{\begingroup \@sanitize@url \@href}%
\providecommand \@href[1]{\@@startlink{#1}\@@href}%
\providecommand \@@href[1]{\endgroup#1\@@endlink}%
\providecommand \@sanitize@url [0]{\catcode `\\12\catcode `\$12\catcode `\&12\catcode `\#12\catcode `\^12\catcode `\_12\catcode `\%12\relax}%
\providecommand \@@startlink[1]{}%
\providecommand \@@endlink[0]{}%
\providecommand \url  [0]{\begingroup\@sanitize@url \@url }%
\providecommand \@url [1]{\endgroup\@href {#1}{\urlprefix }}%
\providecommand \urlprefix  [0]{URL }%
\providecommand \Eprint [0]{\href }%
\providecommand \doibase [0]{https://doi.org/}%
\providecommand \selectlanguage [0]{\@gobble}%
\providecommand \bibinfo  [0]{\@secondoftwo}%
\providecommand \bibfield  [0]{\@secondoftwo}%
\providecommand \translation [1]{[#1]}%
\providecommand \BibitemOpen [0]{}%
\providecommand \bibitemStop [0]{}%
\providecommand \bibitemNoStop [0]{.\EOS\space}%
\providecommand \EOS [0]{\spacefactor3000\relax}%
\providecommand \BibitemShut  [1]{\csname bibitem#1\endcsname}%
\let\auto@bib@innerbib\@empty
%</preamble>
\end{thebibliography}%


\begin{thebibliography}{41}%
%
\bibitem{bennett1993teleporting} 
C. H. Bennett, G. Brassard, C. Cr\'{e}peau, R. Jozsa, A. Peres, and W. K. Wootters, 
Teleporting an unknown quantum state via dual classical and Einstein-Podolsky-Rosen channels,
\textit{Phys. Rev. Lett.}
\href{http://dx.doi.org/10.1103/PhysRevLett.70.1895}{
\textbf{70} 1895
(1993).
}
%
\bibitem{bouwmeester1997experimental} 
D. Bouwmeester, JW. Pan, K. Mattle, M. Eibl, H. Wein, and A. Zeilinger,
Experimental quantum teleportation,
\textit{Nature (London)}
\href{https://doi.org/10.1038/37539}{
\textbf{390} 575
(1997).
}
%
\bibitem{boschi1998experimental} 
D. Boschi, S. Branca, F. De Martini, L. Hardy, and S. Popescu,
Experimental Realization of Teleporting an Unknown Pure Quantum State via Dual Classical and Einstein-Podolsky-Rosen Channels,
\textit{Phys. Rev. Lett.}
\href{https://link.aps.org/doi/10.1103/PhysRevLett.80.1121}{
\textbf{80} 1121
(1998).
}
%
\bibitem{furusawa1998unconditional} 
A. Furusawa, J. L. S\o{}rensen, S. L. Braunstein, C. A. Fuchs, H. J. Kimble, and E. S. Polzik,
Unconditional Quantum Teleportation,
\textit{Science}
\href{https://www.science.org/doi/abs/10.1126/science.282.5389.706}{
\textbf{282} 706
(1998).
}
%
\bibitem{Valivarthi2016quantum} 
R. Valivarthi, M. G. Puigibert, Q. Zhou, G. H. Aguilar, V. B. Verma, F. Marsili, M. D. Shaw, S. W. Nam, D. Oblak, and W. Tittel,
Quantum teleportation across a metropolitan fibre network,
\textit{Nat. Photonics}
\href{http://dx.doi.org/10.1038/nphoton.2016.180}{
\textbf{10} 676
(2016).
}
%
\bibitem{Ren2017ground}
J.-G. Ren, P. Xu, H.-L. Yong \emph{et al}.,
Ground-to-satellite quantum teleportation,
\textit{Nature (London)}
\href{http://dx.doi.org/10.1038/nature23675}{
\textbf{549} 70
(2017).
}
%
\bibitem{Barasinski2019demonstration} 
A. Barsi\'{n}ski, A. \v{C}ernoch, and K. Lemr,
Demonstration of Controlled Quantum Teleportation for Discrete Variables on Linear Optical Devices,
\textit{Phys. Rev. Lett.}
\href{http://dx.doi.org/10.1103/PhysRevLett.122.170501}{
\textbf{122} 170501
(2019).
}
%
\bibitem{zhou2000Methodology} 
X. Zhou, D. W. Leung, and I. L. Chuang,
Methodology for quantum logic gate construction,
\textit{Phys. Rev. A}
\href{http://dx.doi.org/10.1103/PhysRevA.62.052316}{
\textbf{62} 052316
(2000).
}
%
\bibitem{son2001conclusive} 
W. Son, J. Lee, M. S. Kim, and Y.-J. Park
2001
Conclusive teleportation of a d-dimensional unknown state
\textit{Phys. Rev. A}
\href{https://link.aps.org/doi/10.1103/PhysRevA.64.064304}{
\textbf{64} 064304
}
%
\bibitem{Lee2021Quantum} 
S.-W. Lee, D.-G. Im, Y.-H. Kim, H. Nha, and M. S. Kim,
Quantum teleportation is a reversal of quantum measurement,
\textit{Phys. Rev. Res.}
\href{http://dx.doi.org/10.1103/PhysRevResearch.3.033119}{
\textbf{3} 033119
(2021).
}
%
\bibitem{chitambar2023duality} 
E. Chitambar and F. Leditzky,
On the Duality of Teleportation and Dense Coding,
\href{https://doi.org/10.1109/TIT.2023.3331821}{
\textit{IEEE Trans. Inf. Theory}
(2023).
}
%
\bibitem{pan1998experimental} 
J.-W. Pan, D. Bouwmeester, H. Weinfurter, and A. Zeilinger,
Experimental entanglement swapping: entangling photons that never interacted,
\textit{Phys. Rev. Lett.}
\href{https://doi.org/10.1103/PhysRevLett.80.3891}{
\textbf{80} 3891
(1998).
}
%
\bibitem{briegel1998quantum} 
H.-J. Briegel, W. Dur, J. I. Cirac, and P. Zoller,
Quantum repeaters: The role of imperfect local operations in quantum communication,
\textit{Phys. Rev. Lett.}
\href{https://link.aps.org/doi/10.1103/PhysRevLett.81.5932}{
\textbf{81} 5932
(1998).
}
%
\bibitem{lee2019fundamental} 
S.-W. Lee, T. C. Ralph, and H. Jeong,
Fundamental building block for all-optical scalable quantum networks,
\textit{Phys. Rev. A}
\href{https://link.aps.org/doi/10.1103/PhysRevA.100.052303}{
\textbf{100} 052303
(2019).
}
%
\bibitem{gottesman1999demonstarting} 
D. Gottesman and I. L. Chuang,
Demonstrating the viability of universal quantum computation using teleportation and single qubit operations,
\textit{Nature (London)}
\href{https://doi.org/10.1038/46503}{
\textbf{402} 390
(1999).
}
%
\bibitem{pirandola2020advances} 
S. Pirandola, U. L. Andersen, L. Banchi \textit{et al.},
Advances in quantum cryptography,
\textit{Adv. Opt. Photon.}
\href{https://opg.optica.org/aop/abstract.cfm?URI=aop-12-4-1012}{
\textbf{12} 1012
(2020).
}
%
\bibitem{raussendorf2001one} 
R. Raussendorf and H. J. Briegel,
A one-way quantum computer,
\textit{Phys. Rev. Lett.}
\href{https://link.aps.org/doi/10.1103/PhysRevLett.86.5188}{
\textbf{86} 5188
(2001).
}
%
\bibitem{liao2018satellite} 
S.-K. Liao \textit{et al.},
Satellite-Relayed Intercontinental Quantum Network,
\textit{Phys. Rev. Lett.}
\href{https://link.aps.org/doi/10.1103/PhysRevLett.120.030501}{
\textbf{120} 030501
(2018).
}
%q network
\bibitem{daiss2021quantum} 
S. Daiss, S. Langenfeld, S. Welte, E. Distante, P. Thomas, L. Hartung, O. Morin, and G. Rempe,
A quantum-logic gate between distant quantum-network modules,
\textit{Science}
\href{https://www.science.org/doi/abs/10.1126/science.abe3150}{
\textbf{371} 614
(2021).
}

%q network
\bibitem{pompili2021realization} 
M. Pompili, S. L. N. Hermans, S. Baier, H. K. C. Beukers, P. C.Humphreys, R. N. Schouten, R.F.L. Vermeulen, M. J. Tiggelman , L. dos Santos Martins, B. Dirkse, S. Wehner, and R. Hanson,
Realization of a multinode quantum network of remote solid-state qubits,
\textit{Science}
\href{https://www.science.org/doi/abs/10.1126/science.abg1919}{
\textbf{372} 259
(2021).
}
%
\bibitem{kimble2008quantum} 
H. Kimble,
The quantum internet,
\textit{Science}
\href{https://doi.org/10.1038/nature07127}{
\textbf{453} 1023
(2008).
}
%
\bibitem{wehner2018quantum} 
S. Wehner, D. Elkouss, and R. Hanson,
Quantum internet: A vision for the road ahead,
\textit{Science}
\href{https://www.science.org/doi/abs/10.1126/science.aam9288}{
\textbf{362} eaam9288
(2018).
}
%
\bibitem{jonathan1999entanglement} 
D. Jonathan and M. B. Plenio,
Entanglement-Assisted Local Manipulation of Pure Quantum States,
\textit{Phys. Rev. Lett.} 
\href{https://link.aps.org/doi/10.1103/PhysRevLett.83.3566}{
\textbf{83} 3566
(1999).
}
%
\bibitem{lipka2021catalytic}
P. Lipka-Bartosik and P. Skrzypczyk,
Catalytic Quantum Teleportation,
\textit{Phys. Rev. Lett.}
\href{https://link.aps.org/doi/10.1103/PhysRevLett.127.080502}{
\textbf{127} 080502
(2021).
}
%
\bibitem{calsamiglia2001maximum}
J. Calsamiglia and N. L\"{u}tkenhaus,
Maximum efficiency of a linear-optical Bell-state analyzer,
\textit{Appl. Phys. B}
\href{https://doi.org/10.1007/s003400000484}{
\textbf{72} 67-71
(2001).
}
%
\bibitem{knill2001scheme}
E. Knill, R. Laflamme, and G. Milburn,
A scheme for efficient quantum computation with linear optics,
\textit{Nature (London)}
\href{https://doi.org/10.1038/35051009}{
\textbf{409} 46-52
(2001).
}
%
\bibitem{bayerbach2023bell}
M. J. Bayerbach, S. E. D’Aurelio, P. van Loock, and S. Barz,
Bell-state measurement exceeding 50\% success probability with linear optics,
\textit{Sci. Adv.}
\href{https://doi.org/10.1126/sciadv.adf4080}{
\textbf{9} eadf4080
(2023).
}
%
\bibitem{ishizaka2008asymptotic} 
S. Ishizaka and T. Hiroshima,
Asymptotic Teleportation Scheme as a Universal Programmable Quantum Processor,
\textit{Phys. Rev. Lett.}
\href{http://dx.doi.org/10.1103/PhysRevLett.101.240501}{
\textbf{101} 240501
(2008).
}
%
\bibitem{Beigi2011simplified} 
S. Beigi and R. K\"{o}nig,
Simplified instantaneous non-local quantum computation with applications to position-based cryptography,
\textit{New. J. Phys.}
\href{http://dx.doi.org/10.1088/1367-2630/13/9/093036}{
\textbf{13} 093036
(2011).
}
%
\bibitem{may2022complexity} 
A. May,
Complexity and entanglement in non-local computation and holography,
\textit{Quantum}
\href{http://dx.doi.org/10.22331/q-2022-11-28-864}{
\textbf{6} 864
(2022).
}
%
\bibitem{sedlak2019optimal} 
M. Sedl\'{a}k, A. Bisio, and M. Ziman,
Optimal Probabilistic Storage and Retrieval of Unitary Channels,
\textit{Phys. Rev. Lett.}
\href{http://dx.doi.org/10.1103/PhysRevLett.122.170502}{
\textbf{122} 170502
(2019).
}
%
\bibitem{quintino2019reversing} 
M. T. Quintino, Q. Dong, A. Shimbo, A. Soeda, and M. Murao,
Reversing Unknown Quantum Transformations: Universal Quantum Circuit for Inverting General Unitary Operations,
\textit{Phys. Rev. Lett.}
\href{http://dx.doi.org/10.1103/PhysRevLett.123.210502}{
\textbf{123} 210502
(2019).
}
%
\bibitem{buhrman2016quantum} 
H. Buhrman, \L{}. Czekaj, A. Grudka, M. Horodecki, P. Horodecki, M. Markiewicz, F. Speelman, and S. Strelchuk,
Quantum communication complexity advantage implies violation of a Bell inequality,
\textit{Proc. Natl. Acad. Sci. U.S.A.}
\href{http://dx.doi.org/10.1073/pnas.1507647113}{
\textbf{113}, 3191
(2016).
}
%
\bibitem{Pirandola2019fundamental} 
S. Pirandola, R. Laurenza, C. Lupo, and J. L. Pereira,
Fundamental limits to quantum channel discrimination,
\textit{npj Quantum Inf.}
\href{http://dx.doi.org/10.1038/s41534-019-0162-y}{
\textbf{5} 50
(2019).
}
%
\bibitem{Mozrzymas2018optimal} 
M. Mozrzymas, M. Studzi\'{n}ski, S. Strelchuk, and M. Horodecki,
Optimal port-based teleportation,
\textit{New. J. Phys.}
\href{http://dx.doi.org/10.1088/1367-2630/aab8e7}{
\textbf{20} 053006
(2018).
}
%
\bibitem{jeong2020generalization} 
K. Jeong, J. Kim, and S. Lee,
Generalization of port-based teleportation and controlled teleportation, capability
\textit{Phys. Rev. A}
\href{http://dx.doi.org/10.1103/PhysRevA.102.012414}{
\textbf{102} 012414
(2020).
}
%
\bibitem{studzinski2022efficient} 
M. Studzi\'{n}ski, M. Mozrzymas, P. Kopszak, and M. Horodecki,
Efficient Multi Port-Based Teleportation Schemes,
\textit{IEEE Trans. Inf. Theory}
\href{http://dx.doi.org/10.1109/TIT.2022.3187852}{
\textbf{68} 7892
(2022).
}
%
\bibitem{Strelchuk2023minimal} 
S. Strelchuk and M. Studzi\'{n}ski,
Minimal port-based teleportation,
\textit{New. J. Phys.}
\href{http://dx.doi.org/10.1088/1367-2630/acdab4}{
\textbf{25} 063012
(2023).
}
%
\bibitem{grinko2023efficient} 
D. Grinko, A. Burchardt, and M. Ozols,
Efficient quantum circuits for port-based teleportation,
\href{https://doi.org/10.48550/arXiv.2312.03188}{
arXiv:2312.03188
(2023).
}
%
\bibitem{kim2024port} 
H. E. Kim and K. Jeong,
Port-based entanglement teleportation via noisy resource states,
\textit{Phys. Scr.}
\href{https://dx.doi.org/10.1088/1402-4896/ad22c6}{
\textbf{99} 035105
(2024).
}
%
\bibitem{hausladen1994pretty}
P. Hausladen and W. K. Wootters,
A ‘pretty good’measurement for distinguishing quantum states,
\textit{J. Mod. Opt.}
\href{https://www.tandfonline.com/doi/abs/10.1080/09500349414552221}{
\textbf{41} 2385-90
(1994).
}
%
\bibitem{leditzky2022optimality}
F. Leditzky,
Optimality of the pretty good measurement for port-based teleportation,
\textit{Lett. Math. Phys.}
\href{https://doi.org/10.1007/s11005-022-01592-5}{
\textbf{112} 98
(2022).
}
%
\bibitem{peres1991optimal}
A. Peres and W. K. Wootters,
Optimal detection of quantum information,
\textit{Phys. Rev. Lett.}
\href{https://link.aps.org/doi/10.1103/PhysRevLett.66.1119}{
\textbf{66} 1119
(1991).
}
%
\bibitem{badziag2000local} 
P. Badzi\c{a}g, M. Horodecki, P. Horodecki, and R. Horodecki,
Local environment can enhance fidelity of quantum teleportation,
\textit{Phys. Rev. A}
\href{http://dx.doi.org/10.1103/PhysRevA.62.012311}{
\textbf{62} 012311
(2000).
}
%
\bibitem{horodecki1999general} 
M. Horodecki, P. Horodecki, and R. Horodecki,
General teleportation channel, singlet fraction, and quasidistillation,
\textit{Phys. Rev. A}
\href{http://dx.doi.org/10.1103/PhysRevA.60.1888}{
\textbf{60} 1888
(1999).
}
%
\bibitem{ishizaka2009quantum} 
S. Ishizaka and T. Hiroshima,
Quantum teleportation scheme by selecting one of multiple output ports,
\textit{Phys. Rev. A}
\href{http://dx.doi.org/10.1103/PhysRevA.79.042306}{
\textbf{79} 042306
(2009).
}
%
\bibitem{nest2004efficient} 
M. Van den Nest, J. Dehaene, and B. De Moor,
Efficient algorithm to recognize the local Clifford equivalence of graph states,
\textit{Phys. Rev. A}
\href{http://dx.doi.org/10.1103/PhysRevA.70.034302}{
\textbf{70} 034302
(2004).
}
%
\bibitem{studzinski2017port} 
M. Studzi\'{n}ski, S. Strelchuk, M. Mozrzymas, and M. Horodecki,
Port-based teleportation in arbitrary dimension,
\textit{Sci. Rep.}
\href{http://dx.doi.org/10.1038/s41598-017-10051-4}{
\textbf{7} 10871
(2017).
}
%
\bibitem{franson2002high} 
J. D. Franson, M. M. Donegan, M. J. Fitch, B. C. Jacobs, and T. B. Pittman,
High-Fidelity Quantum Logic Operations Using Linear Optical Elements,
\textit{Phys. Rev. Lett.}
\href{https://link.aps.org/doi/10.1103/PhysRevLett.89.137901}{
\textbf{89} 137901
(2002).
}
%
\bibitem{christandl2021asymtotic} 
M. Christandl, F. Leditzky, and C. Majenz \textit{et al.},
Asymptotic Performance of Port-Based Teleportation,
\textit{Commun. Math. Phys.}
\href{https://doi.org/10.1007/s00220-020-03884-0}{
\textbf{381} 379
(2021).
}
%
\end{thebibliography}
\end{document}